\documentclass[preprint,12pt]{aastex}

\usepackage[section]{placeins}

\begin{document}

\title{On Rings and Streams in the Galactic Anti-Center}

\author{
Jing Li\altaffilmark{1,2},
Heidi Jo Newberg\altaffilmark{3},
Jeffrey L. Carlin\altaffilmark{3},
Licai Deng\altaffilmark{1},
Matthew Newby\altaffilmark{3},
Benjamin A. Willett\altaffilmark{3},
Yan Xu\altaffilmark{1},
Zhiquan Luo\altaffilmark{2}
}

\altaffiltext{1}{Key Laboratory of Optical Astronomy, National Astronomical Observatories, Chinese Academy of Sciences, Beijing 100012; lijing@bao.ac.cn}

\altaffiltext{2}{School of Physics and Electronic Information, China West Normal University, Nanchong 637002}

\altaffiltext{3}{Dept. of Physics, Applied Physics and Astronomy, Rensselaer
Polytechnic Institute, Troy, NY 12180, USA; heidi@rpi.edu}

\begin{abstract}
We confirm that there are at least three separate low-latitude over-densities of blue F turnoff stars near the Milky Way anti-center: the Monoceros Ring, the Anti-Center Stream (ACS), and the Eastern Banded Structure (EBS). There might also be a small number of normal thick disk stars at the same location. The ACS is a tilted component that extends to higher Galactic latitude at lower Galactic longitude, 10 kpc from the Sun towards the anti-center. It has a sharp cutoff on the high latitude side. Distance, velocity, and proper motion measurements are consistent with previous orbit fits. The mean metallicity is [Fe/H]$=-0.96 \pm 0.03$, which is lower than the thick disk and Monoceros Ring. The Monoceros Ring is a higher density substructure that is present at $15\arcdeg<b<22\arcdeg$ at all longitudes probed in this survey. The structure likely continues towards lower latitudes. The distances are consistent with a constant distance from the Galactic Center of 17.6 kpc. The mean line-of-sight velocity of the structure is consistent with a thick disk rotation. However, the velocity dispersion of these stars is $\sim 15$ km s$^{-1}$, and the metallicity is [Fe/H]$=-0.80 \pm 0.01$. Both of these quantities are lower than the canonical thick disk. We suggest that this ring structure is likely different from the thick disk, though its association with the disk cannot be definitively ruled out. The Eastern Banded Structure (EBS) is detected primarily photometrically, near $(l,b)=(225\arcdeg,30\arcdeg)$, at a distance of 10.9 kpc from the Sun.
\end{abstract}


\section{Introduction\label{intro}}
An overdensity of photometrically selected F turnoff stars near the Galactic anti-center was first reported by \citet{nyetal02} . These stars were selected to be bluer than the turnoff of the thick disk, as defined by local populations, and have an inferred distance from the Sun of 11-16 kpc. Over-densities of stars were detected in four places, named by Galactic longitude, Galactic latitude, and apparent magnitude in the SDSS g filter: S223+20-119.4, S218+22-19.5, S183+22-19.4, and (with less significance) S200-24-19.8. Two possible explanations for these overdensities were given: a tidal stream, or an ``even thicker disk'' with a scale height of 2 kpc and a scale length around 10 kpc. The detections spanned a range of constellations, so the overdensity was named by a large constellation near the center of the detections ---Monoceros.

In \citet{ynetal03}, evidence was presented that the stars were part of a ring of stars that encircle the Galaxy, and the observations were interpreted as evidence that the stars were from a tidally disrupted satellite galaxy. The argument was based primarily on the small radial velocity dispersion ($\sigma\sim 25$ km s$^{-1}$) of the stars in the structures. This dispersion was similar to the dispersion measured for the Sagittarius (Sgr) dwarf galaxy tidal stream, and much smaller than that measured locally for the thin disk, thick disk or halo. In addition, standard exponential models for the thin and thick disks predicted far fewer star counts in this region of the galaxy than were observed. Note also that the measured metallicity of [Fe/H]$=-1.6$ and turnoff color of $(g-r)_0=0.27$ are similar to the stellar halo and not the thick disk. The evidence in this paper that suggested a thick disk origin for these stars included a stellar distribution that was circularly symmetric about the Galactic center, a circulation velocity that was not significantly different from the thick disk, and a density distribution that fit an exponential (with a vertical scale height of $1.6\pm 0.5$ kpc) as a function of height above the plane. At nearly the same time, \citet{2003MNRAS.340L..21I} announced the ``One ring to encompass them all'', extending the observed Galactic longitudes at which the structure was detected, and favoring its interpretation as a disk perturbation with a vertical scale height of $0.75\pm 0.04$ kpc. 

Following these papers, a flurry of new papers hinted at either a disk or a satellite origin for the ≈``ring'' of stars. \citet{2003ApJ...592L..25H} showed that dwarf galaxy ≈``satellites'' on orbits nearly coplanar to the disk can form ringlike stellar structures as tidal arcs or shells, and hinted that these minor mergers contribute to building the disk component of the Galaxy. \citet{2003PhLB..567....1S}, \citet{2007PhRvD..76b3505N} and \citet{2008PhRvD..78f3508D} argued that the excess of stars on the ``ring'' could be associated with a ringlike overdensity of dark matter (a caustic) that can lead to an overdensity of disk stars. The caustic rings for our galaxy are predicted to be near 40 kpc n$^{-1}$, where n is an integer, so the relevant caustic for observed structure is $n=2$. There might be observational evidence for such rings of dark matter in the rotation curve of the Milky Way \citep{2011JCAP...04..002D}. \citet{2003ApJ...594L.115R} showed that there existed a large group of M giant stars near the anti-center at the same distance as the ``ring'' that extends from $145\arcdeg<l<240\arcdeg$. At $l=270\arcdeg$, the overdensity may be closer to the Galactic center so that it appears to merge with the disk. The M giant stars are not equidistant from the Galactic center, are thicker in some places than others, and appear to have a larger vertical scale height near $l=180\arcdeg$. The presence of M giants indicates a younger, more metal rich population in this structure, supporting the hypothesis that the structure is the remains of the dwarf galaxy with a range of stellar populations rather than a perturbation of the thick disk. \citet{2003ApJ...594L.119C} further measured a metallicity for the M giant stars of [Fe/H]$=-0.4\pm0.3$ and identified four globular clusters of lower mean metallicity that may be associated with the structure, bolstering the claim for a range of metallicities. Their measurement of a line-of-sight, Galactic standard of rest velocity of $V_{gsr}\sim -16$ km s$^{-1}$ at a Galactic longitude of nearly $180\arcdeg$ indicates the structure has a slightly non-circular orbit. \citet{2004ApJ...602L..21F} found 15 old open clusters that appear to form a linear structure that starts near the anti-center, 20 kpc from the center of the Galaxy, and comes closer to the Galactic center as it traverses quadrants III and IV, approaching 6 kpc from the Galactic center when it is on the opposite side of the Galactic center from the Sun. Searches for BHB star substructure in the anti-center were largely unsuccessful \citep{2004ApJ...605L..25K}.

The discussion of the origin of the overdensity in the anti-center became more heated when Martin et al. (2004) announced the discovery of the Canis Major dwarf galaxy, at $(l,b)=(240\arcdeg, -8\arcdeg)$. They argued that this dwarf galaxy was the likely progenitor of the ≈``ring-like structure''. However, both the nature of the Canis major dwarf galaxy and its association with the Monoceros ring, as it came to be known, were controversial. Many articles argued in favor of its identity as a dwarf galaxy \citep{2004AJ....127.3394F,2004MNRAS.354.1263B, 2004MNRAS.348...12M, 2005ApJ...618L..25D, 2005GeoJI.162..448M, 2005nfcd.conf...97M, 2006MNRAS.366..865B, 2007AJ....133.2274B, 2007ApJ...662..259D}. Others claimed the purported dwarf galaxy was an artifact of the Galactic warp or a spiral arm \citep{2004A&A...421L..29M, 2005A&A...442..917C, 2006A&A...451..515M, 2006MNRAS.368L..77M, 2007A&A...472L..47L, 2008A&A...482..777C, 2008MNRAS.390L..54P}. \citet{2006ApJ...640L.147R} claimed that the core of the Canis Major dwarf galaxy was a reddening artifact, and that the low latitude stellar overdensity actually extended over a larger area in the Argo constellation; this larger overdensity was thought to be a dwarf galaxy or disrupting dwarf galaxy, and could have caused the observed warp in HI gas.

The Triangulum-Andromeda (Tri-And) over-density \citep{2004ApJ...615..732R,2004ApJ...615..738M} was discovered in the $100\arcdeg<l<150\arcdeg$, $-20\arcdeg>b>-40\arcdeg$ region, $15-30$ kpc from the Sun. These stars have a narrow velocity dispersion, and a velocity trend that matches the anti-center M giants in \citet{2003ApJ...594L.119C}. However, the Tri-And stars are more distant than the Crane et al. M giants, and have a lower metallicity, leading Rocha-Pinto et al. to wonder whether the Tri-And stars are a more distant wrap of the same tidal stream. \citet{2010ApJ...708.1290C} measure titanium, yttrium, lanthanum, and iron abundances of the Tri-And structure, and conclude that the measurements are consistent with the chemical enrichment of a dwarf galaxy such as Sagittarius, but that it may not be the same dwarf galaxy as the low latitude anti-center structure.

\citet{2005ApJ...626..128P} created a semianalytic n-body simulation of a dwarf galaxy disruption that more or less fit all of the positions in which the anti-center structure had been detected, including the more distant Tri-Andromeda overdensity. The fit was done in sky position, velocity, and distance to previously detected stellar overdensities. One fact to keep in mind here is that most of the ``detections'' of substructure near the Galactic anti-center did not identify a center of the stellar distribution. The observations showed an excess of stars in a particular location, but there could have been a larger excess at a higher or (more likely) lower Galactic latitude. This simulation and fit to the data could not predict with certainty whether or not the reported Canis Major dwarf galaxy was the likely progenitor of the presumed tidal debris.

Positions of additional over-densities of stars (and low latitude positions with non-detections), identified by an unexpected population of stars with a blue turnoff in H-R diagrams were presented by \citet{2005MNRAS.362..475C}, \citet{2006MNRAS.367L..69M}, \citet{2007MNRAS.376..939C}. The over-densities of stars were detected both above and below the Galactic plane. \citet{2008AJ....135.2013C,2010AJ....139.1889C} identified two groups of stars at $(l,b)=(167.1\arcdeg,-34.7\arcdeg)$ and $(122.9\arcdeg,22.4\arcdeg)$ as Monoceros stream candidates by their proper motions. \citet{2010ApJ...714..663D} confirmed the presence of an extra ``ring of stars'' above and below the Galactic plane at $l=94\arcdeg$,$130\arcdeg$, and $150\arcdeg$, but not at $110\arcdeg$. Because the excess stars are detected both above and below the Galactic plane, it is difficult to explain the excess stars as resulting from a disk warp.

\citet{2006ApJ...651L..29G, 2008ApJ...689L.117G} identified a ≈``three-stream complex'' (also dubbed the Anti-Center Stream, ACS) in the anti-center region that was thought to be the result of tidal disruption of a dwarf galaxy of significant size and mass (including globular clusters that are now disrupted). The distance, nearly circular orbit, and main sequence turnoff color of this complex is similar to the ``Monoceros Ring''. However, the orbit of the debris, derived by \citet{2008ApJ...689L.117G} based on sky positions, distances, and radial velocities in two fields overlapping the stream, is tilted with respect to the plane of the Milky Way, so the orbit fit does not pass through either Monoceros or Canis Major. It is therefore identified as a separate structure. One complication that was not identified is that the original detection of the ≈``Monoceros Ring'' in \citet{nyetal02} was not really in Monoceros, either. The \citet{nyetal02} paper identified three overdensities in the anti-center that were thought to be related, describing an area 40 degrees across and spanning many constellations. Monoceros was a large constellation near the center of the three observed overdensities and close to the clearest of them. The overdensity that was most clearly identified as a tidal stream was at $(l,b)=(223\arcdeg,20\arcdeg)$, which is right on the ACS. \citet{2010ApJ...725.2290C} measured three-dimensional kinematics of stream star candidates at $(l,b)=(209\arcdeg,26\arcdeg)$, a position near the edge of the ACS as depicted in \citet{2006ApJ...651L..29G}, and found a velocity that is more parallel to the plane than along the stream direction.

\citet{2006ApJ...651L..29G} also discovered another higher latitude overdensity he called the Eastern Banded Structure (EBS), near $(l,b)=(229\arcdeg,28\arcdeg)$. Although this might have been associated with the ACS, \citet{2011ApJ...738...98G} revisits this structure and concludes the tidal debris is on an eccentric orbit that passes close to both the Galactic center and the Sun.

The debate over whether the Monoceros ring is a general feature of the disk, a disrupted satellite, or a feature of the disk created by a satellite disruption is still far from over. \citet{2008ApJ...676L..21Y} show that a high eccentricity ≈``fly-by'' of a dwarf galaxy could form dynamically cold rings of stars around the Milky Way. This suggestion seems inconsistent with the later measurements of the titanium, yttrium, and lanthanum abundances by \citet{2010ApJ...708.1290C}, which show that 21 anti-center M giant stars have elemental abundances similar to the Sgr dwarf tidal tails, and not similar to Milky Way stars of similar overall metallicity. Note, however, that the Sgr dwarf tidal stream is populated by many BHB stars \citep{2000ApJ...540..825Y}, while the low latitude stream is practically devoid of these stars \citep{2010AJ....140.1850B}, so the two stellar populations clearly differ at low metallicity. Since the \citet{2010ApJ...708.1290C} stars are spread over a large part of the sky, one also wonders whether they are all drawn from the same structure. \citet{2011ApJ...726...47S} show deep imaging of the anti-center region at $(l,b)=(180\arcdeg,21\arcdeg)$ and $(180\arcdeg,25\arcdeg)$ that shows a very narrow main sequence at 9 kpc from the Sun. They conclude it is not consistent with recent Galactic models that include flares and warps. Further \citet{2011ApJ...726...47S} conclude,``in the absence of an `ad hoc' abrupt disk cutoff, the photometric signature of the Mon ring cannot be explained by any smooth variation of the Galactic disk structure.'' \citet{2011MNRAS.414L...1M} point out that the Sgr dwarf galaxy crosses the Galactic plane at approximately the distance of the Monoceros ring, 20 kpc from the Galactic center, and that therefore a satellite progenitor of the ring could have had its orbit circularized by an interaction with the Sgr dwarf galaxy. Alternatively, \citet{2011Natur.477..301P} suggest that ring-like wrappings of spiral arms could be excited in the disk in response to the infall of the Sagittarius dwarf galaxy itself. \citet{2011A&A...527A...6H} have found a disk flare model that fits F and G star counts towards the anti-center. 

In this paper we will use photometry and spectroscopy of stars in the northern hemisphere of the Galaxy, towards the Galactic anti-center, to show that the anti-center contains more than one low-latitude substructure. In section 2, we identify the substructures photometrically. In section 3, we measure the properties of stars that we expect to be thick disk stars, and show that they look very much like the values found elsewhere in the literature. In section 4, we use spectroscopy to measure the properties of the anti-center structures. In section 5, we show that our line-of-sight velocities, distances, and proper motions are consistent with the \citet{2008ApJ...689L.117G} orbit fit to the ACS. In section 6, we discuss the properties of the three identified substructures, using the historical names \citep{2006ApJ...651L..29G}: Monoceros Ring, Anti-Center Stream (ACS), and Eastern Banded Structure (EBS). In section 7, we summarize the conclusions.

The three substructures we identify are at nearly the same distance from the Sun in the anti-center, but have slightly different kinematics, metallicities, and positions in the sky. Previous measurements, explanations, and naming schemes should be re-visited in light of the fact that there are clearly multiple substructures at low latitude near the Galactic anti-center, 20 kpc from the Galactic center. Additionally, we argue that none of these structures are likely to be part of the thick disk, because we do find a few stars with the kinematics and metallicity we expect for thick disk stars. The ``Anti-Center Stream'', ``Monoceros Ring'' and ``Eastern Banded Structure'' all have significantly lower velocity dispersion than the thick disk.

\section{Photometric Identification of the anti-center sub-structures}

The data in this paper are taken from the Sloan Digital Sky Survey Data Release 8, (SDSS DR8; \citep{2011ApJS..193...29A}). We first explore the best color range to select stars that are part of an extended structure near the Galactic anti-center. \citet{nyetal02} reported a turnoff magnitude for this structure of $g_0=19.4$, so we concentrate on stars of about this apparent magnitude. Here, and throughout this paper, the subscript ``0'' indicates the magnitudes have been corrected for extinction, using the \citet{1998ApJ...500..525S} reddening maps, in SDSS DR8. We used the extinction values from \citet[SFD]{1998ApJ...500..525S} that are provided in the SDSS database, without applying the $\sim14\%$ corrections to SFD derived by \citet{2011ApJ...737..103S}. Because nearly all of the data presented here are in relatively extinction-free regions, the difference between the two extinction values should have virtually no effect on our work. For the ACS and EBS regions studied here, $E(B-V) \lesssim 0.1$, which amounts to a difference in $E(g-r)$ between the SFD and \citet{2011ApJ...737..103S} values of $\lesssim 0.002$ magnitudes. This in turn amounts to only a $\sim2\%$ correction to measured distances, which is small compared to the uncertainties in distance measurements reported in this work. (Even at lower latitudes where Monoceros is located, the effect on measured distances is less than $\sim5\%$.) We note also that corrections of $\left| \Delta E(g-r) \right| < 0.005$ have little effect on color-selected samples employed in this work.

\citet{nyetal02} showed that the completeness for $g_0<22$ is not a significant issue in the SDSS image data, and \citet{2011ApJ...743..187N} showed that color errors do not dramatically affect the star counts for stars selected in a narrow color range at magnitudes brighter than $g_0=20.5$, so we do not correct for completeness in our analysis. We also assume throughout this paper that the absolute magnitude of the turnoff stars is $M_g=4.2$, as \citet{2011ApJ...743..187N} showed was typical for low metallicity stellar populations in the Milky Way.

Figure 1 shows the density of anti-center $19<g_0<20$ and $(u-g)_0>0.4$ sources selected from ``STAR'' in SDSS DR8. This ``STAR'' selection ensures that we have only one instance of each stellar object, and that all of our objects are point sources. The $(u-g)_0$ cut removes most quasars from the sample. Stars bluer than $(g-r)_0=0.1$ are evenly distributed on the sky. Stars with $0.1<(g-r)_0<0.2$ show a faint, irregular structure (or possibly more than one structure) with $b<30\arcdeg$ and $190\arcdeg<l<230\arcdeg$, with no data for $l>230\arcdeg$. Stars with $0.2<(g-r)_0<0.3$ show both a significant density near the Galactic plane ($b<22\arcdeg$) and an extended group of stars well above the plane extending ``diagonally'' across the lower-left panel from $(l,b)\sim(150\arcdeg,38\arcdeg)$ to $(l,b)\sim(230\arcdeg,20\arcdeg)$. This feature corresponds to the so-called anti-center Stream (ACS) from \citet{2006ApJ...651L..29G}. In this plot and with low resolution we do not observe the ``banding'' of tributaries found by \citet{2006ApJ...651L..29G}. There is also an overdensity near $(l,b)=(230\arcdeg,25\arcdeg)$ that was identified as the Eastern Banded Structure (EBS) by \citet{2006ApJ...651L..29G}. Figure 1 is consistent with a slightly bluer turnoff for stars at higher Galactic longitude in this substructure. Redder $0.3<(g-r)_0<0.5$ stars at $g_0\sim19.5$ show a concentration of stars at low latitudes, as one would expect for thick disk stars. The very blue turnoff color of the distant, high latitude stars is consistent with the results of previous authors. Throughout this paper, we will concentrate on stars with $0.2<(g-r)_0<0.3$, because this is the color range that appears to show significant high latitude structure in the photometry. 

In Figure 2 we show the magnitude distribution of stars in the bluer color range which is populated primarily by the anti-center substructure, $0.2<(g-r)_0<0.3$, and in the redder color range that includes turnoff stars in the thick disk, $0.3<(g-r)_0<0.5$. The gray scale is kept constant as a function of apparent magnitude, so that the relative number of stars at each magnitude can be compared. Among the bluer $0.2<(g-r)_0<0.3$ stars, the anti-center substructure is clearly dominant at $19<g_0<20$; it is visible but much fainter at $18<g_0<19$, and completely disappears fainter than $g_0=20$. In the bluer $20<g_0<21$ panel, one can see the leading Sagittarius dwarf tidal tail \citep{2003ApJ...599.1082M} arcing down towards the Galactic plane, more distant than the ``Monoceros Ring''.

The right panel of Figure 2 shows stars with the colors of the thick disk turnoff. These stars are present at all magnitude ranges. At bright apparent magnitudes the thick disk extends to higher Galactic latitudes because these brighter stars are closer to the Sun. Additionally, the density of thick disk stars should decrease with increasing distance from the Galactic center (which is closely correlated with distance from the Sun when looking towards the Galactic anti-center at low latitudes). The density of thick-disk stars is decreasing from $g_0=16$ to $g_0=20$. Then, in the $20<g_0<21$ panel, we see both an increase in stars at low-latitude ($b<22\arcdeg$) and stars from the higher-latitude anti-center substructure --- presumably fainter, lower mass main sequence stars. There is also a hint of structure at the same sky location as the Sagittarius leading tidal tail.

If the anti-center substructure arises from a warp and/or a flare of the thick disk, then the warp and/or flare happens to contain stars of a different stellar population than the nearer portions of that thick disk. Let's assume for a moment that the bluer and redder stars are from the same populations. Figure 3 shows the number of stars binned by $(g-r)_0$ color and $g_0$ apparent magnitude. The overdensity at $g_0=19.4$ in the bluest color bin is apparent at $g_0=20.2$ at $0.3<(g-r)_0<0.4$, at $g_0=21$ at $0.4<(g-r)_0<0.5$, and at $g_0=21.8$ at $0.5<(g-r)_0<0.6$. We assume other stellar populations will have similar main sequence slopes. So we compare 1400 stars per bin at$ (g_0,(g-r)_0)=(17,0.25)$ with 4000 stars per bin at $(g_0,(g-r)_0)=(18,0.35)$. There are a factor of three fewer blue stars than red stars in this population we associate with the thick disk. Therefore, if the anti-center substructure has a similar stellar population, then 3600 stars per bin at $(g_0,(g-r)_0)=(19.5,0.25)$ should translate to 10,000 stars per bin at $(g_0,(g-r)_0)=(20.5, 0.35)$. Instead, we only have 5000 stars per bin there, and a much smaller fraction of these redder stars appear to be associated with the narrow substructure. Evidently, the thick disk and anti-center substructure(s) have different stellar populations.

The discontinuity in stellar population is clearly apparent in the color-magnitude diagrams (CMDs) in Figure 4. From Figure 2, we noticed that the anti-center substructure has a higher density component at $b<22\arcdeg$, and a lower density, tilted component at higher latitude. In Figure 4 we show CMDs for lower and higher parts of the sky, with $190\arcdeg<l<200\arcdeg$, and an even lower part of the sky including all available data with $175\arcdeg<l<190\arcdeg$. Both of the lower latitude CMDs have a higher density, but very similar structure to the higher latitude CMD. At $g_0=18$, the turnoff of the thick disk is redder than $(g-r)_0=0.3$. Brighter than $16^{th}$ magnitude, there exist some bluer stars, but it is not clear whether that is a trend or whether it is due to a separate population (for example from the thin disk or other disk substructure) that is mixed with the thick disk in this region. Assuming these bright stars are main sequence turnoff stars, they are closer than 2 kpc from the Sun, and are about a half to one kpc from the Galactic plane. At $g_0=18$, the density of thick disk turnoff stars is clearly diminishing, due to increasing distance from the Galactic center, increasing distance from the Galactic plane, or (most likely) both. 

Fainter than $g_0=18$, we see a very high density main sequence that has a bluer turnoff color than the thick disk, which if anything has been getting redder with increasing apparent magnitude. In the CMD, this happens very discontinuously. The turnoff color and apparent magnitude are identical for the lower latitude and middle latitude portions (in the left and middle panels of Figure 4) of the substructure. The turnoff color and magnitude were determined by selecting apparent magnitude bins of width 0.2 at a variety of magnitudes near the turnoff, and fitting three Gaussians to the histograms of star counts as a function of color. The center of the apparent magnitude range yielding the bluest fit to the peak of the Gaussian is $g_0=19.3$ for the left and center panels of Figure 4. The color of the blue peak is $(g-r)_0=0.310$ for the left panel, and $(g-r)_0=0.306$ for the center panel. These CMDs are very similar to those in Figure~12 of \citet{nyetal02}, which depicts data near $(l,b)=(223\arcdeg, 20\arcdeg)$, and similar low latitude CMDs near the anti-center made by many authors since then. 

The higher latitude turnoff (right panel in Fig. 4) appears to be very slightly bluer than the middle and lower stellar populations. To measure the turnoff magnitude and color, we selected stars in 0.2 magnitude wide slices from Figure 4. A histogram of the $(g-r)_0$ colors in this narrow magnitude range was fit with three Gaussians : one represented the turnoff of the substructure, one represented the red M dwarfs, and the third very wide Gaussian more or less fit the range of main sequence stars in the middle color range. This resulted in a very crude fit to the data, but since the bluest Gaussian seemed to fit the turnoff stars well using this scheme, it was used to measure the color of the turnoff as a function of apparent magnitude. These apparent magnitude slices were shifted brighter and fainter in 0.1 magnitude increments until the bluest of the three Gaussian fits (the one corresponding to the turnoff of the substructure) was the bluest of the fits at any magnitude. The apparent magnitude of the center of the bluest slice was adopted as the turnoff magnitude, and the color of the bluest fit was adopted as the turnoff color. The results for the three panels of Figure 4 were: $(g_0, (g-r)_0)=(19.2, 0.310), (19.2, 0.306)$ and $(19.4, 0.300)$ for panels 1, 2,and 3, respectively. The ``Anti-Center Stream'' is indeed slightly bluer and farther away than the ``Monoceros Ring''.

\section{The properties of the thick disk}

We wish to compare the velocities and metallicities of the stars in the identified anti-center substructure with stars that are in the thick disk. We start by examining the line-of-sight, Galactic standard of rest velocities ($V_{gsr}$) for stars with spectra in SDSS DR8 that have $140\arcdeg<l<240\arcdeg$, $17\arcdeg<b<22\arcdeg$, $19<g_0<20$, and $0.4<(g-r)_0<0.6$ in Figure 5. We used the ``bright normalization'' in Figure 3 of \citet{2008ApJ...673..864J} to estimate the absolute magnitude of thick disk stars of this color to be $M_g=5.7$. At $g_0=19.5$, the implied distance to these stars is 6 kpc. This is the distance used to calculate the expected velocity of the thick disk in this figure. Note that the velocities and metallicities agree with our expectations for thick disk stars \citep{allen,gwk89,s93,cb00,setal03,cetal10,cetal11}.

We next study the metallicity distribution of stars with $140\arcdeg<l<240\arcdeg$, $17\arcdeg<b<22\arcdeg$, $19<g_0<20$, and which have velocities within 50 km s$^{-1}$ of the expected velocity for thick disk stars, in different color ranges. Figure 4 guides us in understanding from which component these stars are likely to be drawn. Figure 6 shows the metallicity distribution for $0.2<(g-r)_0<0.3$ (low latitude substructure), $0.3<(g-r)_0<0.4$ (may include a combination of the low latitude substructure and thick disk), $0.4<(g-r)_0<0.5$ (mostly thick disk), and $0.5<(g-r)_0<0.6$ (thick disk).

The bluest stars, which are part of the low latitude substructure, have a metallicity near [Fe/H]$=-0.8$. Stars with colors $0.4<(g-r)_0<0.6$ have a metallicity near [Fe/H]$=-0.6$, significantly higher than the low latitude substructure. In between, at $0.3<(g-r)_0<0.4$, the stars are plausibly a combination of thick disk stars with a mean of [Fe/H]$=-0.6$ and substructure stars which have a mean metallicity of [Fe/H]$=-0.8$. We do not believe that this transitional color range indicates that there is a gradient in the metallicity of the thick disk, because the CMD in Figure 4 shows a discontinuous transition in density at this color, and the color box includes parts of each distribution. In all color ranges there is a tail of much lower metallicity stars which are stars from the stellar halo that happen to have line-of-sight velocities close to those of the thick disk.

In Figure 7 we show the line-of-sight velocity dispersion of the thick disk as measured from spectra of anti-center stars with $140\arcdeg < l < 240\arcdeg$, $17\arcdeg<b<22\arcdeg$, $19<g_0<20$, and $0.4<(g-r)_0<0.6$ and [Fe/H]$ > -1.0$. As discussed earlier, these stars are about 6 kpc from the Sun, or 14 kpc from the center of the Milky Way, somewhat closer to the Galactic center than the anti-center substructure at about 20 kpc from the Galactic center. For each star, we subtracted the expected line-of-sight velocity for a thick disk star at its position from the measured $V_{gsr}$. We fit a single Gaussian centered at 0.0 km s$^{-1}$ to the residuals, and derived a velocity dispersion for the thick disk of $34.6 \pm 2.0$ km s$^{-1}$. The only other fit parameter was the square-root of the amplitude, which was found to be $7.3\pm 0.3$ counts$^{1/2}$. 

\section{Spectroscopy of the anti-center structure}

In this section we select spectra from SDSS DR8 in the anti-center region. Figure 8 shows a density map of all SDSS DR8 stellar spectra. The density of spectra is substantially higher where SEGUE and SEGUE-2 plates were observed. Spectroscopic plates have a radius of 1.49\arcdeg and take 640 simultaneous spectra, each with wavelength coverage $3800-9200$ \AA~and spectral resolution of $R \sim1800$. SEGUE and SEGUE-2 generally obtained one faint plate and one bright plate at each selected location. The target selection methods were quite complex and varied with time, so it can be challenging to use the data for statistical studies. However, we will be selecting a small range of color and magnitude near the faint end of the spectroscopic survey, so the overall selection criteria will primarily affect only the number of spectra on each plate. In Section 2 we used the photometry to study the relative densities of stars, and in this section we will use the spectra to measure the velocities and metallicities of the stars in the substructure(s).

We studied the properties of the anti-center structures in 34 positions in the sky, identified in Figure 8 and tabulated in Table 1. Because the sets of plates could potentially be sampling different structures at different latitudes, we separated them into ``higher'', ``middle'', and ``lower'' parts based on Galactic latitude. The lower part consists of regions with centers between $5\arcdeg<b<18\arcdeg$, the middle regions have centers between $18\arcdeg<b<28\arcdeg$, and the higher regions are those at $28\arcdeg<b<40\arcdeg$. The designation of each plate to one of these groups is described in Figure 8 and Table 1. Table 1 lists in degrees the geometric center of each selected region, the width of each region in Galactic longitude and latitude, and the average position of the spectra in that region that have $19<g_0<20$ and $0.2<(g-r)_0<0.3$. Also listed is the number of spectra with $19<g_0<20$ and $0.2<(g-r)_0<0.3$. 

The SEGUE Stellar Parameter Pipeline (SSPP; \citet{2008AJ....136.2022L}) processes the wavelength and flux-calibrated spectra generated by the standard SDSS spectroscopic reduction pipeline \citep{2002AJ....123..485S}, and estimates stellar effective temperature, surface gravity, metallicity and heliocentric radial velocity. For each star, we calculated the line-of-sight, Galactic standard of rest values from the heliocentric radial velocities (RV) using the formula:
\begin{equation}
V_{gsr}=RV + 10\cos l\cos b + 225.2\sin l \cos b + 7.2 \sin b {\rm km s}^{-1}
\end{equation}

\noindent which removes the contribution of the 220 km s$^{-1}$ rotation velocity at the Solar circle as well as the Solar peculiar motion (relative to the Local Standard of Rest) of $(U, V, W)_\sun= (10.0, 5.2, 7.2)$ km s$^{-1}$ \citep{1998MNRAS.298..387D}. 

In the Section 2, we showed that the best magnitude and color range to select the anti-center substructure is $0.2<(g-r)_0<0.3$ and $19<g_0<20$. In addition to these criteria, we removed bad data from the SDSS catalogs by selecting only spectra with $elodierverr>0$ (if radial velocities are not measured then $elodierverr$ is zero), $|[Fe/H]|<10$ dex (outside this range the metallicity is clearly incorrect or unmeasured), and that were not missing a magnitude measurement in g or r (resulting in $|(g-r)_0|>10$). This left us with 1300 spectra to analyze. In this paper all metallicities were obtained from the fehspec field in the sppParams table of SDSS DR8. fehspec is the adopted value of the metallicity, using a weighted average of other spectroscopic methods. This does not include photometric methods of measuring metallicity.

Figure 9 shows the SDSS DR8 [Fe/H] vs. line-of-sight, Galactic standard of rest velocity for all of the stars in our 34 selected regions, separated by latitude ranges to show, from left to right, the higher, middle, and lower-part data. The plots clearly show that the low metallicity ([Fe/H]$ < -1.2$) stars in all three latitude ranges have large ($\sim100$ km s$^{-1}$) velocity dispersions expected from a stellar halo population. At higher metallicity, [Fe/H]$>-1.2$, there are apparently a few halo stars but most have a much narrower range of line-of-sight velocities.

In Figure 10, we show the metallicity distribution of blue $[0.2<(g-r)_0<0.3]$, $19.0<g_0<20.0$ stars having velocities within 50 km s$^{-1}$ of the expected thick-disk $V_{gsr}$ along each line of sight for the higher, middle, and lower sets of spectra. We limit the velocity range in order to reduce the number of spheroid stars as much as possible. There are two clear groups: one with low metallicity that matches our expectations for the stellar halo, and a group of stars with metallicities of $-1.2<$ [Fe/H] $<-0.5$. One can see in the lower two panels that the middle and lower spectra, with $b<28\arcdeg$, have a narrow range of metallicities centered at [Fe/H]$=-0.8$. However, the higher spectra ($28\arcdeg<b<40\arcdeg$) have a broader metallicity peak near [Fe/H]$=-1.0$, significantly lower than the lower latitude spectra. 

We fit two Gaussians to the metallicity histogram for the 213 higher Galactic latitude spectra in Figure 10. The stellar halo Gaussian is centered at [Fe/H]$=-1.7$ with a sigma of 0.18. The stream(s) are fit with a Gaussian centered at [Fe/H]$=-0.96\pm 0.03$ and a sigma of 0.31. There were not enough halo stars in the middle and lower panels to fit two Gaussians well, so we calculated the mean and standard deviation of the metallicity from stars around the peak. In the middle part we used metallicities from -1.3 to -0.5, and in the lower part we used metallicities from -1.1 to -0.5. From these, we find mean metallicities of $\langle $[Fe/H]$\rangle = [-0.80, -0.80] \pm [0.01, 0.02]$ for the middle and lower parts, respectively. The sigma of these distributions is found to be 0.20, and 0.19 dex, respectively. Thus, the metallicity distribution of the higher part overlaps those of the middle and lower stars, but with a statistically significant offset in the mean.

We now explore the velocities of the stars in the anti-center substructure. Figure 11 includes nine subsets of our spectroscopic data, separated by three metallicity bins (red lines in Figure 9) and three ranges of Galactic latitude. The low metallicity subsets of the middle and higher latitude data look as we would expect for Galactic halo populations ---a large ($\sim100$ km s$^{-1}$) dispersion centered on $V_{gsr}=0$ km s$^{-1}$. 

In the higher metallicity panels, we show the expected velocities of a thick disk with a circular rotation speed of 170 km s$^{-1}$ and assuming the stars are 10 kpc from the Sun \citet{allen}. We calculated this curve using the following formula: 
\begin{equation}
V_{td}=-(\sin(l)/|\sin(l)|)V_{rot}\times\sin(a)\times\cos(b)
\end{equation}
\begin{equation}
r=\sqrt{[d^2+8^2-2\times 8\times d\times\cos(l)]}
\end{equation}
\begin{equation}
a=\arccos((d^2+r^2-8^2)/(2dr)),
\end{equation}
where 8 kpc is the adopted distance from the Sun to the Galactic center; d is the distance to the observed star (in this case 10 kpc); r is the distance from the Galactic center; ``a'' is the angle between the Sun and the observed star, as viewed from the Galactic center; $V_{rot}$ is the adopted rotation speed of the thick disk (in this case 170 km s$^{-1}$); and $V_{td}$ is the predicted line-of-sight, Galactic standard of rest velocity of the observed star in km s$^{-1}$, assuming it is in the thick disk. 

In the higher latitude, middle metallicity dataset, we see a very narrow velocity dispersion substructure in the range $150\arcdeg<l<190\arcdeg$. The line-of-sight velocities in this range are significantly higher than we expect from the thick disk. In particular, the line-of-sight velocity at $l=180\arcdeg$ is positive, not zero. All axisymmetric disk models predict a zero line-of-sight velocity directly towards the anti-center, independent of rotation speed and distance to the stars, because at that longitude we are looking exactly perpendicular to the circular motion of the disk. At $200\arcdeg<l<210\arcdeg$, there are many fewer spectra, and they appear to be consistent with the velocity of the thick disk, with a dispersion that is believably similar to expectations. At $220\arcdeg<l<230\arcdeg$, the velocity dispersion appears noticeably narrower, but with average line-of-sight velocities consistent with the thick disk.

These observations are quantified in the top two rows of panels in Figure 12. We divided the middle metallicity, higher Galactic latitude data up by numbered region, ignoring regions with fewer than ten spectra. We then fit two Gaussians to the velocity histogram of those data. One of the Gaussians represents the thick disk, with a fixed central $V_{gsr}$ and $\sigma=34.6$ km s$^{-1}$(as derived in Section~3); only the height was allowed to vary. The other Gaussian represents a separate velocity substructure; the height, mean, and sigma of this Gaussian were all allowed to vary. These fits were sometimes sensitive to local minima in chi-squared values, especially when the structure mean and thick disk means were nearly identical, so a Metropolis-Hastings Markov Chain Monte Carlo (MCMC) technique \citep{1986mcm..book.....K} was used to sweep out the parameter space near the fit in order to produce truly minimized best fits. A Hessian matrix, calculated at the best-fit parameters, provided the errors in the parameter fit values. The Hessian matrix was multiplied by 2 and inverted, so that the square root of the diagonal elements gave the parameter variances.

The results of fitting two Gaussians are given in Table 2. The columns are: The name of the region; the expected velocity of the thick disk, in km s$^{-1}$, assuming the stars are 10 kpc from the Sun; the number of stars fit to the thick disk component; the error in the number of stars fit to the thick disk; the mean velocity of the second Gaussian fit; the error in the mean velocity of the second Gaussian fit; the number of stars fit to the second Gaussian; the error in the number of stars fit to the extra Gaussian; the sigma of the second Gaussian; the error in the sigma of the second Gaussian; the mean of the second Gaussian fit minus the expected velocity of the thick disk; the number of velocity standard deviations the mean of the second Gaussian is away from the expected velocity of the thick disk (Dev(V)); the number of stars that were fit; and the color this region is assigned in Figure 13. 

The colors in Figure 13 indicate the type of fit result for the velocities: pink---most of the stars are in a narrower Gaussian that is significantly shifted from the mean of the thick disk (more than 2.5 sigma); black---the stars are mostly consistent with our expectations for the thick disk; green---most of the stars are in a narrower Gaussian that is at the same velocity as is expected for the thick disk; blue---most of the stars are consistent with our expectations for the thick disk, but the distribution is better fit with an additional narrower component that is shifted from the velocity of the thick disk. The two Gaussian fits to H7 and H8 put essentially zero stars in the second Gaussian, and produced random numbers for the velocity profile; these meaningless numbers were deleted from the table to avoid confusion. M13 could have been included in the pink category rather than the green category; there are so many spectra in this region that the formal errors on the velocity are tiny, so even though the velocity offset is less than 6 km s$^{-1}$, it is 2.7 sigma from the thick disk velocity. Since the systematic errors can be as large as 5 km s$^{-1}$, we included this region in the green category, meaning the velocity is not significantly different from our expectations for the thick disk. M4 was also a borderline region. We decided to color it black (consistent with thick disk), because although the fit put ten stars in a second Gaussian, the error in the number of stars in the second Gaussian was larger than that. Region L8 had only 11 spectra in our sample (barely enough to be included in Figure 12). However, only five of those stars were fit in the two Gaussian fit. Because the number of stars was so low and the formal fit was not good, this region was moved to the ``not enough data to fit'' category. 

Higher regions H1, H3, H4, H5, and H6; and middle regions M7 and M9, show a narrow ($\sigma<20$ km s$^{-1}$) velocity peak that is significantly higher in the mean than the expectations for the thick disk (shown by the green line in Figure 12). Of the 20 regions with at least 10 spectra in our color-magnitude selection, only two have velocity dispersions that are well fit by a model thick disk (p-value $>$0.7 in Table 3). Regions H7 and H8 contain very few spectra of which a large fraction are fit by the model thick disk. Regions H10, and H11 show narrow peaks that are quite close to the expected mean velocity of the thick disk, but have $\sigma<20$ km s$^{-1}$ -- much narrower than the set of spectra in plates H7 and H8.

This is consistent with observations of the photometric data, as shown in Figure 13. There are apparently separate high latitude over-densities probed by higher regions H1, H3, H4, H5, and H6; and regions H10 and H11. There is no over-density apparent in regions H7 and H8. 

At low latitude, where the photometric data show a relatively large number of stars in our color-magnitude selection box, we find primarily stars that have a narrow velocity dispersion, centered on the velocity expected for circular rotation with the thick disk. L3, L4, M13 and M14 all have sigmas under 20 km s$^{-1}$. M13 in particular has a sigma of $18\pm 3$ km s$^{-1}$, which is six standard deviations away from the expected thick disk sigma of 34.6 km s$^{-1}$. If the stars in this low latitude over-density are part of a warp or flare of the thick disk, then the thick disk must not only have a lower metallicity at that distance, but it also would need a significantly lower velocity dispersion.

In the low latitude panels of Figure 13, we see a small clump of stars with velocity near $V_{gsr}=-90$ km s$^{-1}$ and Galactic longitude near $l=203\arcdeg$. We isolated a small tight group of four stars in the intermediate metallicity plot that are within a degree of $(l,b)=(203\arcdeg, 8\arcdeg)$, have metallicities within $0.1$ dex of $-0.61$, and have a mean velocity of $V_{gsr}=-90$ km s$^{-1}$. The mean apparent magnitude is $g_0=19.2$, with a very small range. The implied distance to these stars is 10 kpc. The angle from the Sagittarius dwarf galaxy, as measured along the stream, is $\Lambda_\sun=179\arcdeg$. These stars are likely members of the Sagittarius dwarf leading tidal tail, since the position is directly in line with the tidal tail at higher latitude, and the line-of-sight velocity closely matches the \citet{2010ApJ...714..229L} $N$-body simulation of the Sagittarius dwarf tidal disruption in a triaxial Milky Way potential. These stars are more similar to each other in metallicity and velocity than expected for a dwarf galaxy stream, suggesting these might be a substructure (for example a remnant of a globular cluster) within the stream. The model predicts a distance of 15 kpc rather than 10 kpc, which is worrying though not definitively a contradiction due to the very small sample of F turnoff stars. These stars might, for example, be part of a tight substructure within the tidal debris stream. Although the absolute magnitude of $M_g=4.2$ which was used to calculate the distance to these stars is a reasonable estimate for Sagittarius \citep{nyetal02}, it is not necessarily a good estimate for a small clump of stars, especially a clump with a relatively high metallicity.

Regions M3, M4, M7, M10, and M11 are in a transition region. One can see from the data in Figure 12 that they arguably include both a narrow population at the thick disk velocity and an additional narrow population at a slightly higher velocity. When that distribution is fit by a thick disk plus one narrower Gaussian, it is possible to get results that have thick disk plus extra higher velocity narrow component (pink); all thick disk (black); or thick disk plus a narrow component at the same velocity (blue); depending on the fraction in each extra population. The fact that M3 and M7 have a component that has the same velocity as the structure at higher Galactic latitude leads us to believe that this higher structure extends at least down to $b=26\arcdeg$. The high latitude side of the structure appears to stop abruptly, but the low latitude side is populated by similar stars.

To more rigorously separate the thick disk and this structure, we fit a single free Gaussian to the velocity data in these regions. The results are in Table 3. To further strengthen our argument, we performed a Kolmogorov-Smirnov (K-S) test on each region, with the null hypothesis being that the each data set was derived from the thick disk velocity distribution. The p-values are also listed in Table 3. Not surprisingly, the two regions most consistent with being drawn from the thick disk are H7 and H8. These distributions might actually be drawn from the thick disk. The only regions that might sensibly be drawn from a distribution with the velocity of the thick disk and a sigma on the velocity distribution of 34.6 km s$^{-1}$ are H6, H10, H11, M3, M4, M10, M11, and L4. All ten other regions have p-values less than 0.1, implying, for each region, that there is less than a $10\%$ chance of discovering the observed distribution, provided that only thick disk stars are present. These ten regions are then significantly distinct from a region that is dominated by the thick disk, and therefore likely contain structures that are significantly different from the thick disk. 

\section{The Orbit of the Anti-Center Stream}

In this section we compare our measurements of the position, distance, and velocities of stars in the ACS with the orbit calculated by \citet{2008ApJ...689L.117G}. Figure 16 shows the sky positions where we have both spectra and measurements of turnoff star densities. Red points show the positions of stars with the metallicities and velocities that we expect for ACS stars. We used the average latitude and longitude of regions H3, H4, H5, and H6, as listed in Table 1, as the stream positions. Since M9 is significantly below the northern edge of the ACS, and the likely ACS stars are near the north end of the spectroscopic region, we used the sky position $(l,b)=(203\arcdeg,27\arcdeg)$, which was estimated from the photometry. Regions M13/14 similarly seemed a bit below the top edge of the ACS, though it is much harder to tell due to confusion with the Monoceros Ring. We used a sky position of $(l,b)=(225\arcdeg, 20\arcdeg)$ for the last datapoint. We used the $V_{gsr}$ and the error in the $V_{gsr}$ for each region as tabulated in Table 2. For region M13/14, we used average values of V$_{gsr} = -51\pm 9.6$ km s$^{-1}$.

The green outlined regions in Figure 16 were selected to include ACS stars and be loosely centered on interesting spectroscopic regions. We found that stars from the leading tidal tail of the Sagittarius dwarf spheroidal galaxy were pulling our distance measurements farther away as Galactic longitude approached $200^\circ$, so we also selected a higher latitude region as a reference (green dashed line in Figure 16).

Figure 17 shows the magnitude distribution as a function of Galactic longitude for the reference region, the high latitude portion of the ACS, and the ACS with the reference region subtracted. Figure 18 shows histograms with a Gaussian fit to the peak apparent magnitude for Galactic longitudes between $160\arcdeg$ and $200\arcdeg$, with the reference region (containing Sagittarius debris) subtracted. The peak apparent magnitudes for the regions centered at $162.5\arcdeg, 167.5\arcdeg, 172.5\arcdeg, 177.5\arcdeg, 182.5\arcdeg, 187.5\arcdeg, 192.5\arcdeg$ and $197.5\arcdeg$, are 19.46, 19.38, 19.42, 19.31, 19.38, 19.32, 19.38, and 19.72, respectively. Note that the ACS stays at nearly a constant distance from the Sun. The last histogram, with stars in the range $195\arcdeg<l<200\arcdeg$, is substantially shifted and broadened, and we suspect it is contaminated with Sagittarius dwarf tidal debris, so we do not use this distance when comparing with the ACS orbit.

Figure 19 shows a comparison of our measurements of position, distance, and line-of-sight velocity with the orbit fit by \citet{2008ApJ...689L.117G}. The distances were determined by assuming that the average turnoff star has an absolute magnitude $M_g=4.2$. Our data is a very good fit, and supports the idea that the ACS is a co-rotating tidal debris stream.

We also checked whether the SDSS proper motions were consistent with the \citet{2008ApJ...689L.117G} orbit fit. The mean predicted proper motion of the orbit between $160\arcdeg < l < 190\arcdeg$ is $(\mu_\alpha \cos\delta, \mu_\delta) = (0.72,-0.30)$ mas yr$^{-1}$. For comparison, we extracted proper motions from SDSS DR7 for all stars between $160\arcdeg < l < 190\arcdeg$, $29\arcdeg < b < 38\arcdeg$, $19 < g_0< 20$, and $0.2 < (g-r)_0 < 0.3$. Histograms of these proper motions are seen in Figure 20. The median proper motion from SDSS data is denoted
by a solid vertical line, with the expected motion in a non-rotating halo given as a dashed line, disk-like rotation of 220 km s$^{-1}$ given as a dotted line, and the predictions of the \citet{2008ApJ...689L.117G} orbit as dot-dashed lines. We note that the
$\mu_\delta$ proper motions for ACS turnoff candidates from SDSS are between the predictions for the halo and thin-disk populations. Note also the apparent double peak in this dimension of the proper motions. We expect about half of the stars in the sample to be halo stars and half to be members of the ACS. For example, note in the top two panels of Figure 11 that in the $160^\circ<l<190^\circ$ region, about half of the stars have the velocity and metallicities of halo stars and half have the velocity and metallicity of the ACS; very few appear to be disk stars. In Figure 21 we show the measured direction of the ACS, as determined from SDSS proper motions. The proper motion is just along the overdensity of stars, as we would expect. Since half of the stars in the sample are halo stars (stationary on average), the magnitude of the tangential velocity is artificially small but the direction is unchanged.

We now re-visit a previous, confusing measurement of the ACS proper motion. \citet{2010ApJ...725.2290C} measured the space velocity of stars near $(l,b)=(209\arcdeg,26\arcdeg)$ that had colors and magnitudes consistent with membership in the ACS. Their result showed that the selected stars appear to move in a direction inclined by about 30 degrees to the visible stream, roughly along Galactic latitude. The authors suggested that because the field of view where their data were taken is on the periphery of the visible stream, they may be seeing the peculiar motion of a sub-component about the center of mass of the progenitor system. Here we note that the ACS has a sharp density cutoff at its high-latitude edge; the field in the Carlin et al. study appears to lie in a region of lower density slightly beyond this cutoff (see Figure 16). Thus, since we see the lower-latitude structure extending up to almost b$\sim27$ degrees, it may be that the Carlin et al. study sampled part of the lower-latitude substructure (at its extension to higher latitudes) rather than the higher-latitude, inclined, stream-like structure. This assertion is bolstered by the fact that the 3-D motion measured by \citet{2010ApJ...725.2290C} of stars selected from a clear stellar overdensity in the color-magnitude diagram is nearly parallel to the Milky Way disk.

To assess the possible contribution of Monoceros Ring-like stars to the Carlin et al. result, we selected only the stars among their 31 reported stream candidates consistent with our findings for the substructure(s) in this region. Namely, we selected only stars bluer than $(g-r)_0=0.4$, fainter than $g_0=18.7$ (i.e., right on the narrow main sequence turnoff locus of the anti-center structure), and with metallicity between $-1.5<$ [Fe/H] $<-0.5$. From the remaining 15 stars, we find weighted mean proper motions of $(\mu_{\alpha} \cos \delta, \mu_\delta)=(-0.55, -0.58)\pm (0.40, 0.33)$ mas/yr [compare to the original result of $(\mu_{\alpha} \cos \delta, \mu_\delta)=(-1.20, -0.78)\pm (0.34, 0.36)$mas/yr. When converted to proper motions along Galactic coordinates (i.e., $l$ and $b$) in a Galactocentric frame (i.e., removing the contribution of the Sun's 220 km s$^{-1}$circular velocity), these become $(\mu_l\cos b, \mu_b)\arcmin=(-3.86, 0.54)\pm (0.34, 0.39)$ mas yr$^{-1}$ [compared to $(\mu_l\cos b, \mu_b)\arcmin=(-4.11, -0.02)\pm (0.36, 0.34)$ mas yr$^{-1}$ in \citet{2010ApJ...725.2290C}]. This exercise brings the measured proper motions to an orientation inclined by $\sim15$ degrees (or $\sim2.5$ sigma) to the visible stream (Figure~21), though also still consistent with disk-like rotation at just above the 1-sigma level. We note that the lower substructure in our study has been found to have essentially disk-like motion among blue turnoff stars, so that the \citet{2010ApJ...725.2290C} study may have been sampling stars from the ñlower substructureî identified here, rather than the higher feature. The 15 stars in our selection from the Carlin et al. data yield a weighted mean line-of-sight velocity of V$_{gsr} = -12$ km s$^{-1}$, and mean metallicity of [Fe/H]$= -1.0$.

\section{The Nature of the anti-center Substructures}

Along our line of sight at $b=20\arcdeg$ (see center panel of Figure 4), the number of blue stars in the thick disk is steadily decreasing with distance until suddenly a group of stars with a different metallicity, different color turnoff, similar mean velocity, narrower velocity dispersion, and narrow distance range appear at of $3-6$ kpc above the disk and 20 kpc from the Galactic center. The anti-center substructure appears to include (at least) three components with different properties: a low latitude, higher density substructure (the Monoceros Ring); a prominent higher latitude substructure that is tilted with respect to the disk (the ACS); and a smaller substructure (EBS) that is near $(l,b)=(220\arcdeg,30\arcdeg)$.

The measured properties of thick disk stars 6 kpc from the Sun towards the Galactic anti-center, in this same SDSS DR8 dataset are: rotation speed consistent with 170 km s$^{-1}$ (this was assumed based on the compilation of \citet{allen}, not fit, but is a good match to the data), velocity dispersion of $34.6\pm2.0$ km s$^{-1}$, metallicity approximately -0.6, including stars with metallicity $-1.0<$[Fe/H]$<-0.1$.

\subsection{Anti-Center Stream (ACS)}

The primary higher latitude component (ACS) does not have constant Galactic latitude as a function of Galactic longitude. It extends above $b=35\arcdeg$ at $l=160\arcdeg$, and is below $b=25\arcdeg$ at $l=220\arcdeg$. The high latitude edge appears quite sharp, and the density appears to decrease towards lower Galactic longitudes. It does not disappear, however, since we still find stars that are kinematically associated with this structure in spectroscopic regions M7, and probably also M3, which are almost $10\arcdeg$ below the top edge of the structure. This component has a mean metallicity of [Fe/H]$=-1.0$, a very narrow velocity dispersion ($\sigma=15$ km s$^{-1}$), and a line-of-sight velocity that is higher than expected for the thick disk. In particular, the line-of-sight $V_{gsr}$ is positive at $l=180\arcdeg$. In all axially symmetric potential models for the Milky Way, disk structures will have a line-of-sight velocity in this direction of zero, since on average the stars are on circular orbits and at $l=180\arcdeg$ we are looking at that circle in the radial direction. We show that the line-of-sight velocities, distances, and proper motions of ACS stars are consistent with the orbit fit of \citet{2008ApJ...689L.117G}. 

\subsection{The Monoceros Ring Component}

The lower latitude component has a much higher density of stars, is present at $b<22\arcdeg$ at all longitudes probed in this survey, has a metallicity [Fe/H]$=-0.8$, has a velocity dispersion of $\sigma\sim15$ km s$^{-1}$ (as calculated from regions M14, L3, and L4), has a mean line-of-sight velocity that is consistent with the 170 km s$^{-1}$ rotation speed of the thick disk, and spans a Galactic longitude range of at least $175\arcdeg<l<230\arcdeg$. 

From the data presented in Figure 14, we compute the distance to the Monoceros Ring along four low-latitude lines of sight. We selected four low latitude regions, outlined by blue lines in Figure 15, and histogrammed the apparent magnitudes of the blue ($0.2<(g-r)_0<0.3$) stars in each line of sight. From the histograms in Figure 14, we measured the peak apparent magnitude, which is tabulated in Table 4. We assume that the mean absolute magnitude of F turnoff stars in low metallicity populations is $M_g=4.2$ \citep{2011ApJ...743..187N}, and calculate the distance to the Monoceros Ring in each direction. From Figure 16 we estimate the mean Galactic coordinates in the selected low latitude regions, and then convert to Galactic (X, Y, Z), assuming the Sun is 8 kpc from the Galactic center. We use the right-handed convention that (X,Y,Z) are zero at the Galactic center, X increases in the direction from the Sun to the Galactic center, Y increases in the direction of Galactic rotation, and Z is positive in the North Galactic Cap. The Monoceros ring stays at an approximately constant cylindrical distance from the Galactic center, consistent with the finding that the mean velocity is the same as a (circularly symmetric) thick disk. Because the Monoceros feature is not clearly distinct in velocity and does not have a well-defined position, we choose not to try to fit an orbit to it -- more information on both the kinematics and the position of the highest density portion of Monoceros will be needed to derive an orbit for this structure.

Because the Monoceros ring is circularly symmetric and co-rotating with the disk, it has been difficult to separate it from the disk populations. In \citet{ynetal03} it was argued that we could separate it from the thick disk because it had a much lower velocity dispersion and a larger scale height. Larger scale heights, we thought, would require larger velocity dispersions. However, it was later pointed out to us that it was possible to have a larger scale height and lower velocity dispersion if the surface density in the Galactic plane decreased at 20 kpc from the Galactic center.

In this paper, our argument that the Monoceros Ring is different from the thick disk depends on two lines of reasoning. The first point is that the Monoceros Ring stars at 20 kpc from the Galactic center are clearly drawn from a different population than the thick disk stars at about 14 kpc from the Galactic center. Our data shows a discontinuity in the number of stars detected along our line of sight, and the metallicity and turnoff color of those stars. The Monoceros Ring is lower metallicity, has a bluer turnoff, and is much more densely populated than the expected thick disk at that distance and height above the plane. If the Monoceros Ring is a warp or flare of the thick disk, its stellar populations change and then flare between 14 and 20 kpc from the Galactic center. The disk would have to have a significant gradient in metallicity with distance from the Galactic center, the velocity dispersion would have to decrease by half, and then the disk would have to flare up (or warp) and end abruptly. 

The second point is that we see evidence that there are some stars with the expected kinematics of the thick disk that are observed in the same color and apparent magnitude range (and therefore presumed distance from the Galactic center) as the Monoceros ring. The stars in regions H7, H8 in Figure 12 and Table 2 have a distribution expected for the thick disk, and no observed over-density in the number counts of stars. The stellar distribution is completely consistent with being drawn from a thick disk population, as shown by a K-S test. In addition, there are indications of a low fraction of stars with thick disk kinematics in most of the other anti-center regions with spectra in the $19<g_0<20$ range. Figure 14 supports the idea that the spectra in regions H7, H8 have some higher metallicity stars, in keeping with the thick disk hypothesis, and that there is a second population of disk-like stars in high latitude regions where we see the ACS. It would be difficult to explain two sets of stars from the same Galactic component, in the same location, that have different velocity dispersions.

Many previous authors who have studied disk star counts towards the Galactic anti-center have concluded that the disk cuts off at 12-15 kpc from the Galactic center \citep{1992ApJ...400L..25R, 1996ApJ...468..663F, 1996A&A...313L..21R, 2009A&A...495..819R, 2010MNRAS.402..713S}. A recent study by \citet{2011ApJ...733L..43M} examines five directions with Galactic longitudes near 60\arcdeg, 95\arcdeg, 295\arcdeg, 314\arcdeg, and 335\arcdeg, and find that the disk cuts off around $13-14$ kpc from the Galactic center in all directions. Although we find a few stars that could be members of the thick disk at 20 kpc from the Sun in higher regions 7 and 8, our data are consistent with a significant drop in density around 14 kpc from the Galactic center near the anti-center.

If there are indeed stars with thick disk kinematics, including a larger velocity dispersion, at the same distance as we see the Monoceros Ring, then it becomes harder to associate the Monoceros Ring with the thick disk.

We would also like to point out that the stellar density substructure in the panel of Figure 1 with $0.2<(g-r)_0<0.3$ and $19<g_0<20$ is not at all similar to the density structure in a similar region of Figure 7 from \citet{2005ApJ...626..128P}. The $N$-body simulation in this figure does not have a stellar density that increases towards the Galactic plane. Near the anti-center, the stars are spread fairly evenly in the range $0\arcdeg<b<40\arcdeg$. There is no enhanced density at $b<22\arcdeg$ as we see in the data. Although one might be able to adjust the \citet{2005ApJ...626..128P} simulation of tidal disruption in the Galactic anti-center to fit the ACS, the Monoceros ring is not explained by this model. 

\subsection{Eastern Banded Structure (EBS)}

We see an over-density of stars in the same sky location and at about the same distance as \citet{2006ApJ...651L..29G} identified the Eastern Banded Structure (EBS). In Figure~15, we showed that the peak apparent magnitude of the turnoff of this structure is $g_0=19.88$, which corresponds to a distance of 13.67 kpc from the Sun, assuming an absolute magnitude of $M_{g}=4.2$. This is somewhat larger than the distance of $9.7\pm0.9$ kpc found by \citet{2011ApJ...738...98G} and the $\sim10$ kpc found by \citet{2009ApJ...703.2177S} to their detections of velocity substructures apparently within the EBS labeled ``B-7/PCI-8/PCII-20" and ``B-8/PCI-9/PCII-21". However, as noted previously, the starcount peak in Figure~15 is broad compared to what one expects from the CMD in that region. We thus need to assess whether additional (sub-)structures are ``contaminating" our EBS distance-measurement sample.

The upper panel of Figure~22 shows a color-magnitude Hess diagram for the $2.5\arcdeg$ wide region centered on the EBS seen as a red box in Figure~16, spanning $220\arcdeg < l < 230\arcdeg$ between $(-0.15 l + 62.1\arcdeg) < b < (-0.15 l + 64.6\arcdeg)$. In this figure it is clear that there is a turnoff at $g_0 \sim 19.4$, which corresponds to a distance of $\sim10.9$ kpc from the Sun, assuming an absolute magnitude of $M_{g}=4.2$. A comparison region of the same size but at $5\arcdeg$ higher Galactic latitude was also selected. The Hess diagram of this region, seen in the lower panel of Figure~22, does not show the clear main sequence seen in the EBS region, but it does show a feature at fainter ($g_0 \gtrsim 19.5$) magnitudes. 
Upon close examination, this ``extra" feature is apparent in both panels of Figure~22 with the highest density at $g_0 \sim 20.3-20.5$, corresponding to a distance of $\sim16.6-18.2$ kpc (again assuming $M_{g}=4.2$).
This additional peak is likely responsible for broadening the EBS peak in Figure~15.
We explore this in Figure~23, which shows histograms of $g_0$ magnitudes of F turnoff stars having $0.2 < (g-r)_0 < 0.3, (u-g)_0 > 0.4$ for the same $2.5\arcdeg$ regions used to generate the ``EBS" and ``background" CMDs in Figure~22. The EBS region shows two apparent peaks at $g_0 \sim 19.5$ and $g_0 \sim 20.4$ on top of the Milky Way halo distribution. The background region, displaced $5\arcdeg$ to higher latitude, shows the ``extra" peak at $g_0 \sim 20.4$, but does not have the peak at brighter magnitudes that we believe is due to the EBS. This becomes clear when we subtract the off-stream field from the EBS region's histogram (bottom panel in Figure~23) -- the residuals are a rather narrow feature that we find to be well fit with a Gaussian at $g_0 = 19.40^{+0.14}_{-0.13}$. This corresponds to a distance to the EBS of $10.96^{+0.73}_{-0.66}$ kpc, assuming $M_g = 4.2$.

The mean line-of-sight velocities of EBS stars are nearly identical to expectations for the thick disk at that distance, but the velocity dispersion is somewhat narrower. There are few enough stars that a K--S test does not rule out the possibility that these stars are selected from a thick disk population, though the best fit is lower dispersion and there is clearly an over-density here in the photometry. From Figure~14, we see that the metallicity of the EBS is about [Fe/H]$\sim-0.8$, and clearly lower than the ACS.

\section {Conclusions}

In this paper we study the anti-center substructure, originally discovered by \citet{nyetal02} at about 20 kpc from the Galactic center. We confirm that this substructure is composed of at least three substructures: the Monoceros Ring, the Anti-Center Stream, and the Eastern Banded Structure \citep{2006ApJ...651L..29G}. We also suggest that there might be a small number of normal thick disk stars at the same Galactocentric radius towards the anti-center, thus bolstering the case that the anti-center substructures are not extensions of the thick disk at large distances from the Galactic center.

The properties of the substructures are: 

{\it Thick Disk:} The thick disk, at 6 kpc from the Sun towards the anti-center, has line-of-sight velocities consistent with a 170 km s$^{-1}$ rotation speed, and a velocity dispersion of $\sigma=34.6\pm 2.0$ km s$^{-1}$. The metallicity distribution of these stars peaks at about [Fe/H] $=-0.6$.

Two of the regions with spectra, H7 and H8 at $(l,b)\sim(205\arcdeg,30\arcdeg)$, have $19<$ g$_0<20$ F turnoff stars with mean line-of-sight velocity and velocity dispersions that are identical to the thick disk. Figure 14 (top panel) shows that although there are a few disk-like stars with [Fe/H] $\sim-1.0$, there are also disk-like stars with [Fe/H] $\sim-0.6$.

Spectra from regions at similar latitudes, apparent magnitudes, and colors, but which cover sky directions in which one of the three anti-center substructures is located, do not have a high fraction of stars with thick disk metallicities (though there are a few). Most of the spectroscopic regions have a few stars that are consistent with thick disk dispersion, and the line-of-sight velocity distribution is better fit assuming there are some thick disk stars in addition to the narrower substructure stars (Figure 12).

If there are a few thick disk stars at the same location as the substructure, it strengthens the case that these substructures are not the result of a warp or flare of the thick disk.

{\it Anti-Center Stream (ACS):} This is a tilted component that extends to higher Galactic latitude at lower Galactic longitude towards the anti-center. It appears to have a sharp cutoff on the high latitude side. Although the density of the stars appears highest at the high latitude edge, the structure appears to include stars at least down to $b=25\arcdeg$, even when the high latitude edge is over $b=35\arcdeg$.

The line-of-sight velocities, distances, and proper motions along the high latitude edge of the substructure are consistent with the orbit fit in \citet{2008ApJ...689L.117G}. The anti-center stream is likely to be an artifact of tidal disruption of a dwarf galaxy, though $N$-body simulations should be run to show that the observed density structure can be fit. 

The mean metallicity is [Fe/H] $=-0.96\pm0.03$, which is lower than the thick disk and Monoceros Ring.

{\it Monoceros Ring:} This is a higher density substructure that is present at$15\arcdeg<b<22\arcdeg$ at all longitudes probed in this survey. The structure likely continues towards lower latitudes. The high density of stars at lower latitudes is not consistent with the \citet{2005ApJ...626..128P} model. The distances are consistent with a constant cylindrical distance from the Galactic center of 17.6 kpc, assuming the Sun is 8 kpc from the Galactic center. The mean line-of-sight velocity of the structure is consistent with a rotation speed of 170 km s$^{-1}$, similar to the thick disk rotation speed and direction. However, the velocity dispersion of these stars is $\sim15$ km s$^{-1}$, and the metallicity is [Fe/H]$=-0.80\pm0.01$. Both of these quantities are lower than the canonical thick disk.

We suggest that this ring structure is likely different from the thick disk, though its association with the disk cannot be definitively ruled out. All indications from our data and from previous authors are that the thick disk stellar density drops off quickly farther than 14 kpc from the center. If the Monoceros Ring is a thick disk structure, then the data implies the thick disk changes properties dramatically after 14 kpc: the scale height increases while the dispersion decreases; the metallicity decreases and the turnoff becomes bluer; and the disk suddenly disappears after flaring up. If the small number of stars with a broader velocity distribution do in fact belong to the canonical thick disk population and are at the same distance as the Monoceros Ring, then it is more difficult to claim the Monoceros Ring is a part of the thick disk.

{\it Eastern Banded Structure (EBS):} This structure is detected primarily photometrically, near $(l,b)=(225\arcdeg,30\arcdeg)$, at a distance of 10.9 kpc from the Sun. The small number of spectra available in this region of the sky have line-of-sight velocity similar that expected for the thick disk, but again with a narrower best fit velocity dispersion ($<20$ km s$^{-1}$). The metallicity of the EBS is about [Fe/H]$=-0.8$.

In addition to these main conclusions, we serendipitously discovered a few stars that are most likely part of the Sagittarius leading tidal tail. These stars at $(l,b)=(203\arcdeg, 8\arcdeg)$ have V$_{gsr}=-90$ km s$^{-1}$, [Fe/H]$=-0.6$, and have apparent magnitude g$_0=19.2$. The sky position and velocity of these stars is consistent with the \citet{2010ApJ...714..229L} triaxial halo model.

Clearly, a larger and more complete spectroscopic survey of the anti-center is warranted, to further clarify the nature and origin of these substructures. We would like 3D kinematics (proper motions and radial velocities) as well as higher resolution chemical abundance analysis. Upcoming surveys including The Guoshoujing Telescope (LAMOST), BigBOSS, HERMES, APOGEE, and Gaia will be quite useful for these studies. 

We thank Dana I. Casetti for suggesting that we check the SDSS proper motions, and the anonymous referee for helpful suggestions that greatly improved the manuscript. Funding for this research was provided by National Science Foundation grants AST 09-37523 and AST 10-09670, and NSFC grants 10973015 and 11061120454.

Funding for SDSS-III has been provided by the Alfred P. Sloan Foundation, the Participating Institutions, the National Science Foundation, and the U.S. Department of Energy Office of Science.The SDSS-III web site is http://www.sdss3.org/.

SDSS-III is managed by the Astrophysical Research Consortium for the Participating Institutions of the SDSS-III Collaboration including the University of Arizona, the Brazilian Participation Group, Brookhaven National Laboratory, University of Cambridge, Carnegie Mellon University, University of Florida, the French Participation Group, the German Participation Group, Harvard University, the Instituto de Astrofisica de Canarias, the Michigan State/Notre Dame/JINA Participation Group, Johns Hopkins University, Lawrence Berkeley National Laboratory, Max Planck Institute for Astrophysics, New Mexico State University, New York University, Ohio State University, Pennsylvania State University, University of Portsmouth, Princeton University, the Spanish Participation Group, University of Tokyo, University of Utah, Vanderbilt University, University of Virginia, University of Washington, and Yale University.

\clearpage
\begin{deluxetable}{lccccccc}
\tabletypesize{\scriptsize}
\tablewidth{0pt}
\tablecolumns{8}
\tablecaption{Positions of regions with spectra}
\tablehead{
\colhead{Region name} & \colhead{Center } & \colhead{Center} & \colhead{Width in} & \colhead{Height in} & \colhead{Average} & \colhead{Average} & \colhead{\# of faint } \\
\colhead{ } & \colhead{Longitude ($^{\circ}$)} & \colhead{Latitude($^{\circ}$)} & \colhead{longitude ($^{\circ}$)} & \colhead{latitude($^{\circ}$)} & \colhead{longitude ($^{\circ}$)} & \colhead{latitude($^{\circ}$)} & \colhead{blue spectra}} 
\startdata
H 1 & 150.00 &31.70 &4.00 &7.00 &149.9676 &31.2386 &67\\
H 2 & 152.90 &38.40 &4.00 &4.00 &153.1573 &39.0832 &7\\
H 3 & 162.50 &36.50 &4.00 &7.00 &163.0226 &37.1451 &48\\
H 4	&173.00	&36.40	&4.00	&6.00	&172.6470	&36.0995	&52\\
H 5	&183.00	&33.70	&6.00	&5.00	&182.9574	&33.5342	&69\\
H 6	&191.00	&32.30	&6.00	&6.00	&190.8146	&32.3594	&73\\
H 7	&200.50	&33.00	&7.00	&6.00	&200.1001	&33.0672	&50\\
H 8	&210.00	&29.60	&6.00	&6.00	&209.8005	&28.7036	&55\\
H 9	&217.00	&31.70	&4.00	&4.00	&216.3567	&31.5812	&24\\
H 10	&221.00	&28.20	&4.00	&5.70	&221.0135	&27.9826	&34\\
H 11	&225.00	&29.00	&4.00	&3.30	&225.1213	&29.0196	&40\\
M 1	&150.00	&26.00	&4.00	&3.60	&---	& ---	&0\\
M 2	&150.00	&20.00	&4.00	&4.00	&150.2018	&19.9595	&6\\
M 3	&177.00	&26.10	&6.00	&5.00	&177.4108	&26.1866	&64\\
M 4	&182.00	&22.70	&6.00	&3.00	&181.1404	&22.8006	&42\\
M 5	&178.00	&20.20	&4.00	&3.00	&178.2038	&20.0677	&23\\
M 6	&187.00	&20.00	&4.00	&3.00	&186.8184	&19.9398	&17\\
M 7	&192.00	&24.90	&6.00	&5.00	&192.1945	&24.5137	&62\\
M 8	&198.00	&21.60	&4.00	&6.00	&197.9617	&22.0361	&24\\
M 9	&204.00	&25.60	&4.00	&4.00	&203.7196	&25.2022	&15\\
M 10	&202.50	&20.50	&5.00	&5.00	&202.4722	&20.4903	&59\\
M 11	&211.50	&18.70	&3.00	&3.40	&211.4654	&18.6389	&35\\
M 12	&215.00	&19.80	&4.00	&3.40	&215.0576	&19.0822	&14\\
M 13	&223.00	&18.70	&4.00	&4.00	&222.9287	&18.6869	&292\\
M 14	&229.00	&20.10	&4.00	&4.00	&228.8201	&19.8727	&24\\
L 1	&150.00	&15.20	&4.00	&4.00	&149.7269	&15.5047	&3\\
L 2	&178.00	&15.00	&4.00	&3.70	&178.0495	&14.7283	&14\\
L 3	&187.00	&17.00	&4.00	&3.00	&187.1922	&17.1901	&15\\
L 4	&187.00	&13.00	&4.00	&5.00	&187.3288	&13.4486	&17\\
L 5	&187.00	&8.00	&4.00	&4.00	&187.2947	&8.7601	&3\\
L 6	&198.00	&8.00	&4.00	&3.00	&197.9669	&8.0067	&7\\
L 7	&203.00	&16.00	&4.00	&3.50	&202.7882	&16.2208	&16\\
L 8	&203.00	&8.20	&4.00	&4.00	&203.0770	&8.1779	&18\\
L 9	&229.00	&14.10	&4.00	&4.00	&229.0977	&14.2354	&11\
\enddata
\end{deluxetable}

\clearpage
\begin{deluxetable}{lccccccccccccc}
\tabletypesize{\scriptsize}
\rotate
\tablewidth{0pt}
\tablecolumns{14}
\tablecaption{Results of fitting thick disk (td) and substructure (h) Gaussian to the velocity distribution in each region. $V_{td}$ is the expected velocity of the thick disk. $N_{td}$ is the number of stars fit to the thick disk Gaussian. $V_h$,$N_h$,and $\sigma_h$ are the parameters for a Gaussian fit to the rest of the stars. The error in each quantity X is given by $\delta X$. $\Delta V$ is the difference between the velocity of the substructure and the expected velocity of the thick disk. $Dev(v)$ is the difference between the thick disk and substructure Gaussians. N is the number of stars observed in the given region.}
\tablehead{
\colhead{Region} & \colhead{$V_{td}$} & \colhead{$N_{td}$} & \colhead {$\delta N_{td}$}&\colhead {$ V_h$} &\colhead{$\delta V_h $}&\colhead{$ N_h$}&\colhead {$\delta N_h$}&\colhead{$\sigma_h$}&\colhead{$\delta \sigma_h$}&\colhead{$\Delta V$}&\colhead{Dev(V)}&\colhead{$ N $}&\colhead{}\\ 
\colhead{}&\colhead{km s$^{-1}$}& \colhead{}&\colhead{}&\colhead{km s$^{-1}$}&\colhead{km s$^{-1}$}& \colhead{}&\colhead{}&\colhead{km s$^{-1}$}&\colhead{km s$^{-1}$}&\colhead{km s$^{-1}$}&\colhead{}&\colhead{}&\colhead{}}
\startdata
H 1 & $33.26$ & $9.15$ & $9.02$ & $61.91$ & $4.32$ & $15.39$ & $11.07$ & $10.32$ & $6.68$ & $28.7$ & $6.6$ & $24.0$ & Pink \\
H 3 & $18.48$ & $3.82$ & $11.56$ & $45.92$ & $11.12$ & $14.45$ & $8.21 $ & $17.35$ & $9.81$ & $27.4$ & $2.5$ & $23.0$ & Pink\\
H 4 & $ 7.43$ & $3.73$ & $8.77 $ & $42.85$ & $5.38 $ & $22.87$ & $ 8.95$ & $14.02$ & $5.59$ & $35.4$ & $6.6$ & $27.0$ & Pink\\
H 5 & $-3.29$ & $6.66$ & $14.01$ & $24.0 $ & $ 9.6 $ & $25.09$ & $9.89 $ & $17.72$ & $8.87$ & $27.3$ & $2.8$ & $36.0$ & Pink\\
H 6 & $-12.24$ & $5.6 $ & $11.14$ & $10.56$ & $6.69 $ & $21.28$ & $9.31$ & $15.28$ & $5.76$ & $22.8$ & $3.4$ & $36.0$ & Pink\\
H 7 & $-22.55$ & $10.38$ & $6.97$ & $---$ & $ ---$ & $ 0.0 $ & $ 0.0 $ & $--- $ & $ --- $ & $--- $ & $--- $ & $13.0$ & Black\\
H 8 & $-33.99$ & $15.81$ & $7.59$ & $---$ & $ ---$ & $ 0.0 $ & $ 0.0 $ & $ --- $ & $ --- $ & $--- $ & $--- $ & $17.0$ & Black\\
H 10& $-46.6 $ & $ 2.03$ & $31.41$& $-52.53$& $12.77$ & $13.25$ & $15.58$ & $19.13$ & $21.56$& $-5.9$ & $-0.5$ & $17.0$ & Green\\
H 11& $-50.52$ & $2.73$ & $15.91$& $-52.87$& $10.06$ & $9.69$ & $ 8.31$ & $14.25$ & $13.0$ & $-2.4$ & $-0.2$ & $16.0$ & Green\\
M 3 & $3.55 $ & $36.06$ & $15.3 $& $31.36$ & $13.95$ & $9.97$ & $8.65$ & $13.2 $ & $12.41$& $27.8$ & $2.0 $ & $44.0$ & Blue\\
M 4 & $-2.43 $ & $13.3 $ & $11.17$& $-17.26$& $9.50$ & $9.76$ & $11.72$ & $9.72 $ & $10.78$& $-14.8$& $1.6 $ & $25.0$ & Black\\
M 7 & $-14.33$ & $14.8 $ & $12.69$& $ 10.73$& $6.81$ & $18.21$ & $8.55 $ & $12.9 $ & $6.59 $& $25.1 $& $3.7 $ & $38.0$ & Pink\\
M 9 & $-28.33$& $0.0 $ & $1.23$& $-3.05$ & $8.82$ & $11.77$ & $7.43 $ & $ 14.74$& $ 9.33$& $25.3$ & $2.9$ & $13.0$ & Pink\\
M 10 & $-27.61 $& $ 16.92$& $2.46$& $-9.13$ & $16.3$ & $6.54$ & $3.64 $ & $12.84$ & $ 5.11$& $18.5 $& $1.1$ & $27.0$ & Blue\\
M 11 & $-38.83$ & $ 11.9 $& $5.98$& $-4.05$ & $33.55$& $3.31$ & $6.49 $ & $10.36$ & $39.01$& $34.8 $& $1.0 $ & $16.0 $& Blue\\
M 13 & $-52.41$ & $45.54 $& $40.0$& $ -46.5$ & $2.18$ & $161.07$ & $24.99$ & $18.27$ & $2.69$ & $5.9$ & $2.7$ & $209.0$& Green\\
M 14 & $-58.77$ & $0.0$ & $0.16$& $-54.6 $ & $6.82$ & $15.23$ & $7.15$ &$14.04$ & $5.55$ & $4.17$& $0.6$ & $ 16.0$& Green\\
L 3 & $-8.82$& $2.78$& $ 14.5$& $-11.93 $& $10.84$& $10.34$&$8.33$& $ 13.93$& $9.7$& $-3.1$& $-0.3$& $13.0$& Green\\
L 4 & $-8.99$& $0.0$ & $1.45$ & $-12.07$ & $10.68$& $12.21$& $6.74$& $17.44$& $7.85$& $-3.1$& $-0.3$& $13.0$& Green\\
L 8 & $-29.81$& $1.57$& $6.53$& $-31.78$ & $61.38$& $3.34$ & $20.65$&$ 4.34$& $31.56$& $-2.0$& $ 0.0 $& $11.0 $& ---\\
\enddata
\end{deluxetable}

\clearpage
\begin{deluxetable}{lccccccc}
\tabletypesize{\scriptsize}
\rotate
\tablewidth{0pt}
\tablecolumns{8}
\tablecaption{Results of fitting a single Gaussian to velocity histograms in each of the regions. The mean of the fit Gaussian is $V_{\rm fit}\pm \delta V_{\rm fit}$ and the sigma of the fit Gaussian is $\sigma \pm \delta\sigma$. The p-value is a measure of how well the distribution matches the default expectation for the velocity distribution of the thick disk. The ``color'' indicates the previous classification based on the results of fitting two Gaussians to the velocity histogram. For example, regions for which the two Gaussian fit preferred to put all of the stars in the thick disk (black) have a P-value indicating a high probability of being drawn from a single Gaussian.}
\tablehead{
\colhead{Region} &\colhead{V$_{\rm td}$} & \colhead{V$_{\rm fit}$} & \colhead{$\delta V_{\rm fit}$} & \colhead{$\sigma$} & \colhead{$\delta\sigma$} & \colhead{p value} & \colhead{``color'' }\\
\colhead{ } &\colhead{km s$^{-1}$} & \colhead{km s$^{-1}$} & \colhead{km s$^{-1}$} & \colhead{km s$^{-1}$} & \colhead{km s$^{-1}$} & \colhead{ } & \colhead{ }} 
\startdata
H10	&-46.60	&-52.14	&10.75	&20.26	&10.27	&0.31	&Green\\
H11	&-50.52	&-53.80	&10.45	&17.34	&11.59	&0.45	&Green\\
M13	&-52.41	&-47.19	&1.85	&20.61	&1.65	&$8.4\times10^{-8}$	&Green\\
M14	&-58.77	&-54.53	&6.83	&14.03	&5.53	&0.079	&Green\\
L3	&-8.82	&-15.29	&9.58	&16.98	&8.93	&0.098	&Green\\
L4	&-8.99	&-11.97	&10.68	&17.47	&7.83	&0.28	&Green\\
H7	&-22.55	&-17.85	&21.90	&25.39	&18.59	&0.82	&Black\\
H8	&-33.99	&-41.90	&19.70	&31.48	&16.69	&0.97	&Black\\
M4	&-2.43	&-9.31	&35.62	&19.98	&36.38	&0.41	&Black\\
M3	&3.55	&12.70	&8.94	&32.04	&7.61	&0.25	&Blue\\
M10	&-27.61	&-22.84	&8.81	&21.84	&6.86	&0.54	&Blue\\
M11	&-38.83	&-26.21	&17.35	&28.33	&14.33	&0.20	&Blue\\
H1	&33.26	&50.96	&15.45	&20.39	&15.16	&0.0016	&Pink\\
H3	&18.48	&40.06	&12.41	&19.84	&8.87	&0.046	&Pink\\
H4	&7.42	&37.60	&10.99	&16.72	&6.78	&$2.2\times10^{-7}$	&Pink\\
H5	&-3.29	&17.34	&7.87	&20.74	&5.22	&0.0025	&Pink\\
H6	&-12.24	&7.20	&8.63	&19.96	&8.44	&0.19	&Pink\\
M7	&-14.33	&3.15	&7.86	&23.33	&6.52	&0.032	&Pink\\
M9	&-28.33	&-9.26	&16.83	&16.35	&12.24	&0.0053	&Pink\\
L8	&-29.81	&-42.91	&65.63	&34.77	&44.67	&0.049	&---\\
\enddata
\end{deluxetable}

\clearpage
\begin{deluxetable}{lcccccccc}
\tabletypesize{\scriptsize}
\rotate
\tablewidth{0pt}
\tablecolumns{9}
\tablecaption{Distances to the Monoceros Ring at four Galactic longitudes. The table gives the apparent magnitude of the turnoff of the Monoceros Ring. $D_\Sun$ is the distance from the Sun, assuming M$_g=4.2$ for these turnoff stars. $(l,b)$ are the Galactic coordinates where the Monoceros Ring stars were sampled. ``(X,Y,Z)'' are the Galactocentric coordinates of each Monoceros Ring detection, and R$_{GC}$ is the cylindrical distance of that detection from the Galactic center.}
\tablehead{
\colhead{} & \colhead{$m_{g_0}$} & \colhead{$D_{\sun}$(kpc)} & \colhead {$l\arcdeg$}&\colhead {$b\arcdeg$}&\colhead{X (kpc)}&\colhead{Y (kpc)}&\colhead {Z (kpc)}&\colhead{$R_{GC}$ (kpc)}}
\startdata
LS1	&19.19	&10.0	&178	&16.5	&-17.6	&0.3	&2.8	&17.6\\
LS2	&19.15	&9.8	&188	&16.5	&-17.3	&-1.3	&2.8	&17.3\\
LS3	&19.36	&10.8	&205	&16.5	&-17.4	&-4.4	&3.1	&17.9\\
LS4	&19.47	&11.3	&225	&16.5	&-15.7	&-7.7	&3.2	&17.4\\

\enddata
\end{deluxetable}

\clearpage
\setcounter{page}{1}

\begin{figure}
\centering
\label{skyplot}
\figurenum{1}
\includegraphics[scale=0.65,angle=0]{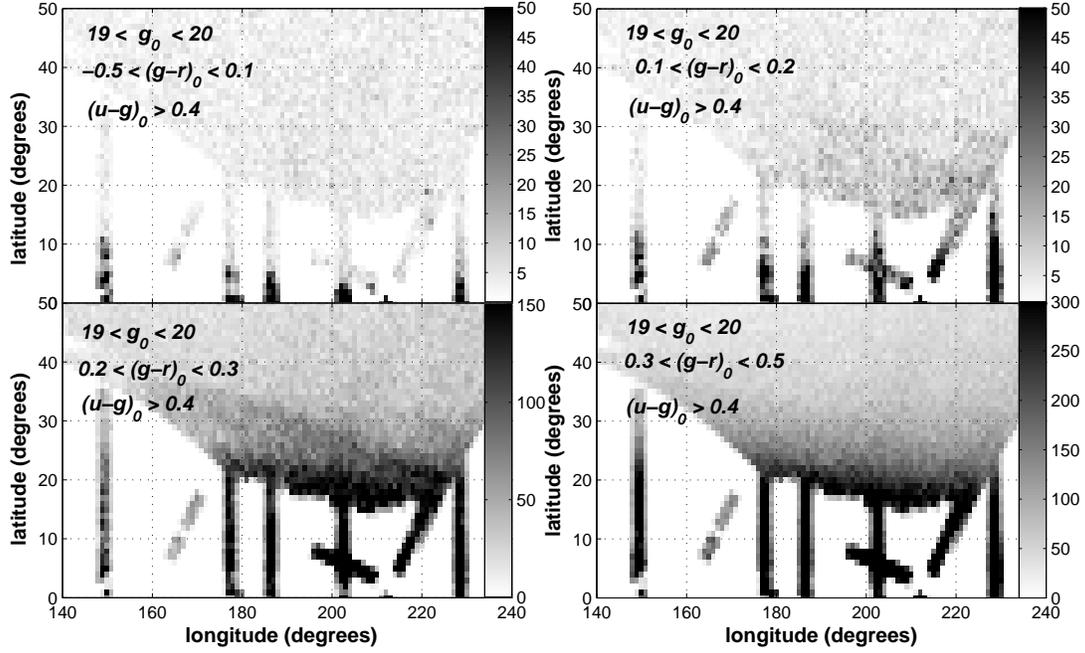}
\caption{Density plot of SDSS DR8 stars with $19<g_0<20$ and $(u-g)_0>0.4$, in different color ranges of $(g-r)_0$. The upper left shows stars with $-0.5<(g-r)_0<0.1$; the upper right panel shows stars with $0.1<(g-r)_0<0.2$; the lower left panel shows stars with $0.2<(g-r)_0<0.3$; and the lower right panel shows stars with $0.3<(g-r)_0<0.5$. The color bar indicates the number of stars in an area of sky covering one degree of Galactic latitude and one degree of Galactic longitude. Notice that there are very few stars bluer than $(g-r)_0=0.3$, since this is bluer than the turnoff for the thick disk and spheroid. From this figure, we tentatively identify three separate substructures, and conclude that the best color range for selecting stars in these anti-center structures is $0.2<(g-r)_0<0.3$.}
\end{figure}

\begin{figure}
\figurenum{2}
\begin{minipage}[c]{0.5\textwidth}
\centering
\includegraphics[width=3in]{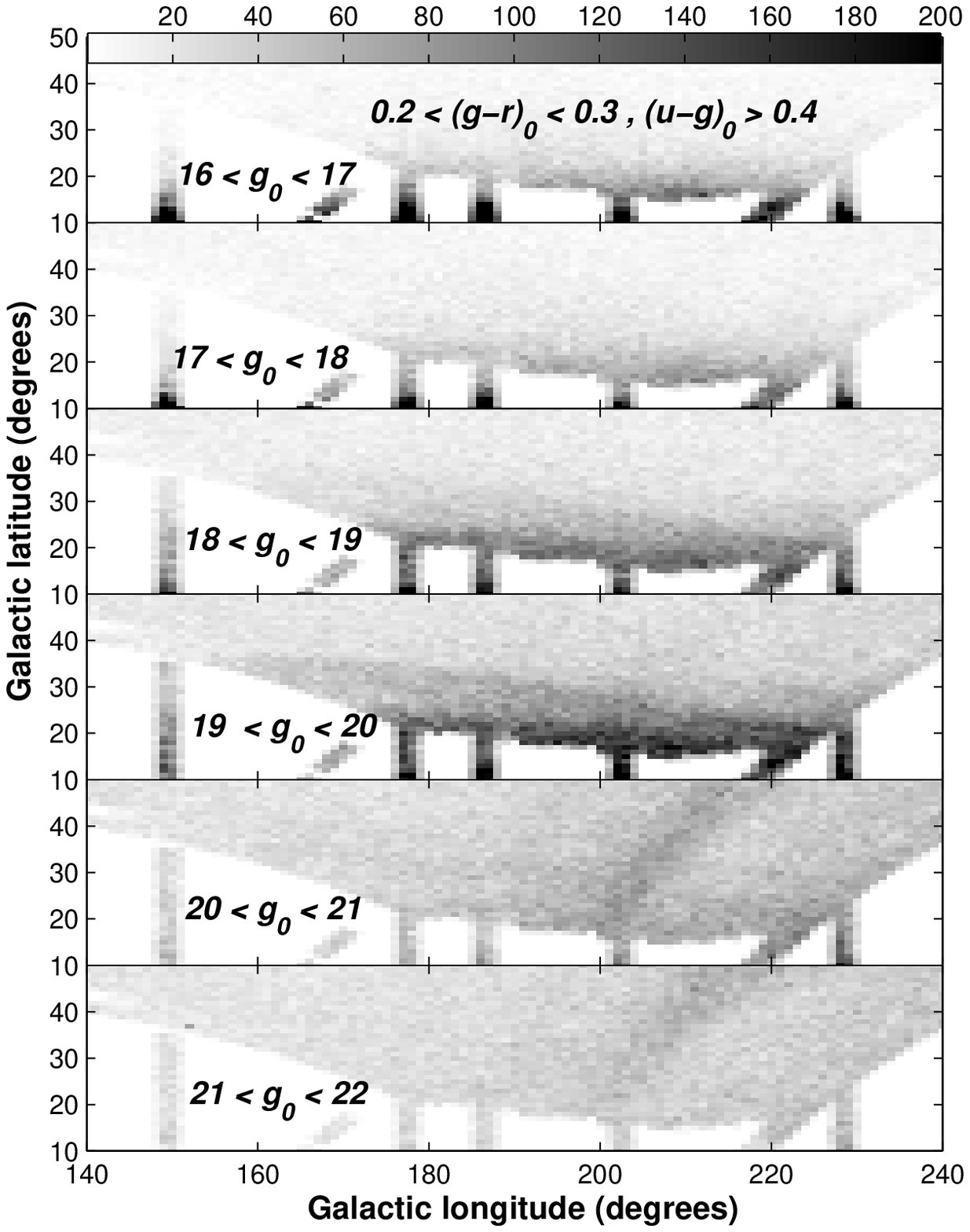}
\end{minipage}
\begin{minipage}[c]{0.5\textwidth}
\centering
\includegraphics[width=3in]{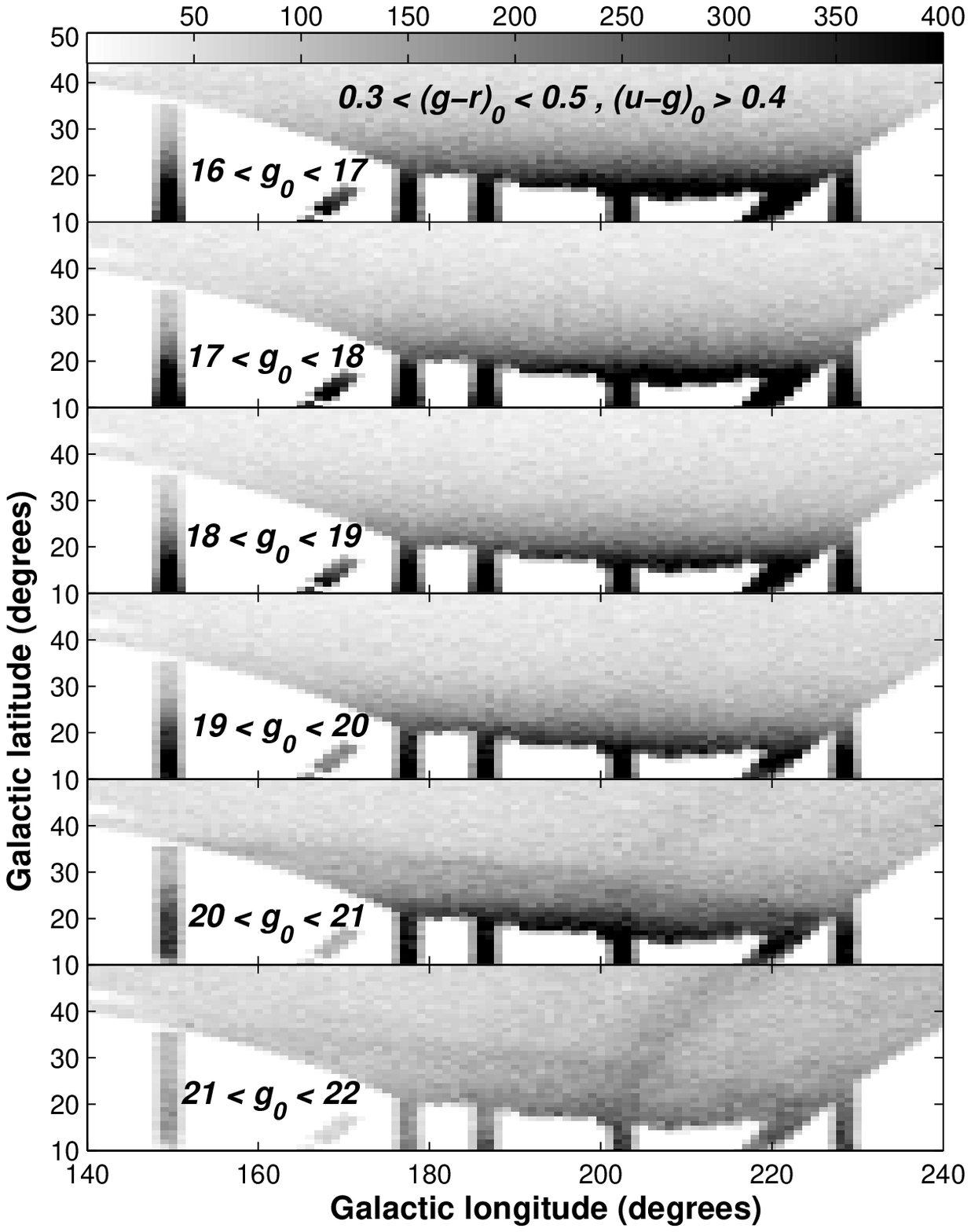}
\end{minipage}

\caption{Magnitude distribution of color-selected stars, with $(u-g)_0>0.4$. The left panel shows stars with colors that favor anti-center substructure. The right panel shows stars with colors that favor the thick disk. Notice that the left panels show a little anti-center substructure at $18<g_0<19$, but most of the ``excess'' stars are at $19<g_0<20$. The substructure appears to be more dense at low latitude ($b<20\arcdeg$) and there is a higher, tilted component that extends up to $b=35\arcdeg$ at $l=180\arcdeg$. In contrast, the thick disk stars are strong at all apparent magnitudes. Stars from the tilted component of the anti-center substructure are apparent, but not dominant, at $20<g_0<21$ in this redder color range. This figure shows that: (1) The anti-center substructures have a bluer turnoff, and are therefore drawn from a different stellar population from the thick disk; (2) the low latitude ``Monoceros Ring'' is at about the same distance as the ``Anti-Center Stream'' (ACS), and unlike the thick disk stars is contained within a very narrow distance range; (3) the thick disk star counts with $0.3<(g-r)_0<0.5$ appear to decrease from $g_0=16$ to $g_0=20$, then increase in the $20<g_0<21$ region, where we see the fainter main sequence stars of the anti-center substructure; and (4) fainter than the anti-center substructure no concentration of stars near the Galactic plane is apparent.}
\end{figure}

\begin{figure}
\centering
\figurenum{3}
\includegraphics[scale=0.6,angle=0]{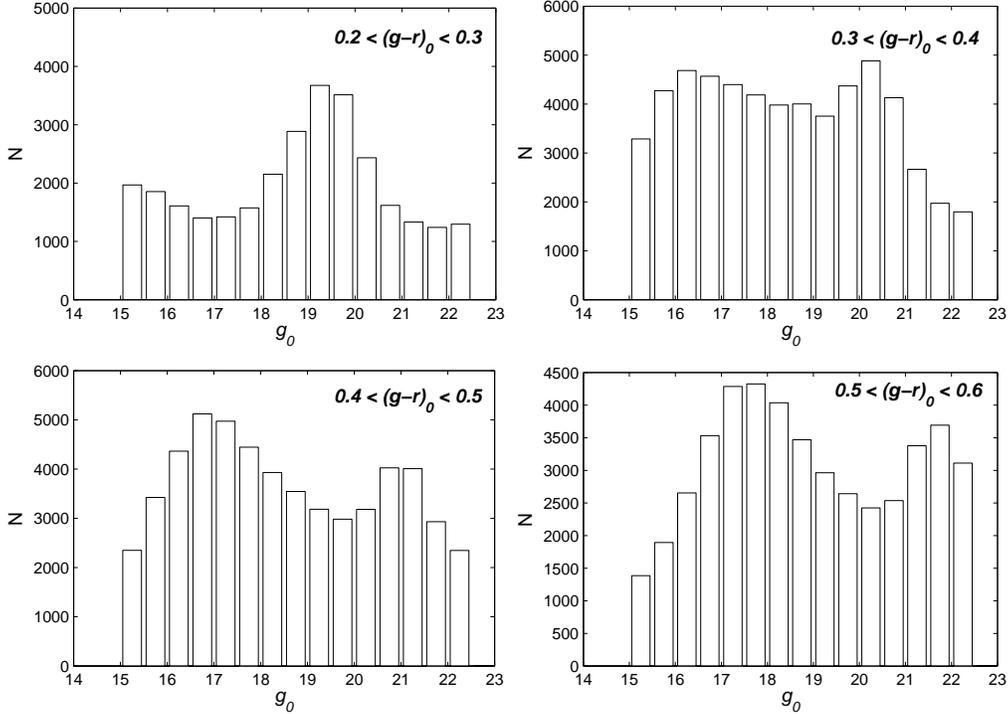}
\caption{Density vs. apparent magnitude in different color ranges. Each panel shows a histogram of the number of stars as a function of $g_0$ apparent magnitude, for different $(g-r)_0$ color ranges [top left: $0.2<(g-r)_0<0.3$; top right: $0.3<(g-r)_0<0.4$; lower left: $0.4<(g-r)_0<0.5$; and lower right: $0.5<(g-r)_0<0.6$]. The stars were selected from the region of the sky given by $210\arcdeg<l<215\arcdeg$, $10\arcdeg<b<30\arcdeg$ that is populated by SDSS DR8 photometric data. The $(u-g)_0>0.4$ color cut has been applied. Although SDSS data are essentially complete for $g_0<22$, stars begin to be scattered significantly in and out of narrow color selection bins fainter than $g_0>20.5$. The apparent magnitude of the anti-center substructure in each panel shifts as the implied absolute magnitude of the stars being sampled increases. In order of increasing $(g-r)_0$ color, the peak is at $g_0$=19.4, 20.2, 21.0, and 21.8. In all but the top left panel, which contains stars mostly blueward of the thick disk turnoff, we see a much broader peak of thick disk stars. At bright magnitudes, the number of stars increases as the volume increases with distance from the Sun. At fainter magnitudes the number of stars decreases as the density of thick disk stars falls off exponentially with distance from the Galactic center and distance from the Galactic plane.}
\end{figure}

\begin{figure}
\centering
\figurenum{4}
\includegraphics[scale=0.6,angle=0]{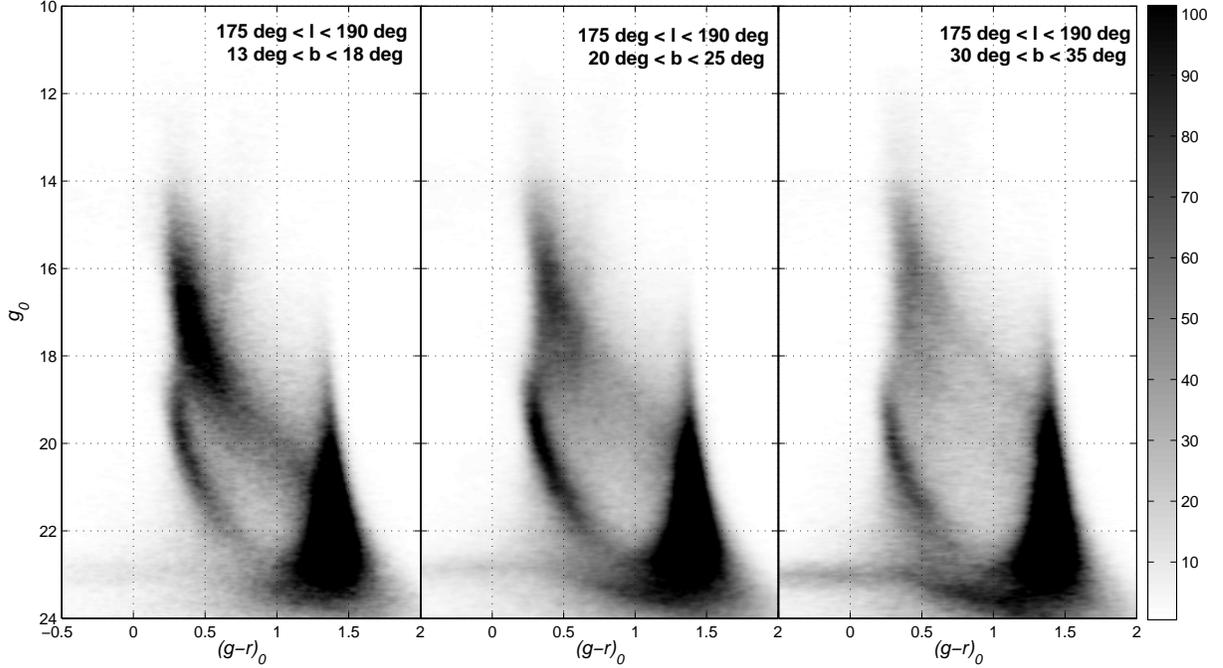}
\caption{Color-Magnitude Hess diagram of lower, middle, and higher latitude substructures. These stars were selected from SDSS DR8 with psfmagerr\_g and psfmagerr\_r both between zero and one, and $(u-g)_0>0.4$. Fainter than $g_0=18$, we see a narrow main sequence with a bluer turnoff than the brighter thick disk stars. The lower (left panel) and middle (center panel) latitude regions of the sky show identical turnoff color and magnitude. The gap between the thick disk stars and the low latitude substructure is even more apparent at $b=15\arcdeg$ than it is at $b=22\arcdeg$. (Note the relative paucity of stars with $(g-r)_0=0.5$ and $g_0=20$, and with $(g-r)_0=0.8$ and $g_0=21.5$ at low latitude.) The higher latitude substructure has a slightly bluer turnoff color and fainter apparent magnitude than those of the lower-latitude features. }
\end{figure}

\begin{figure}
\centering
\figurenum{5}
\includegraphics[scale=0.6,angle=0]{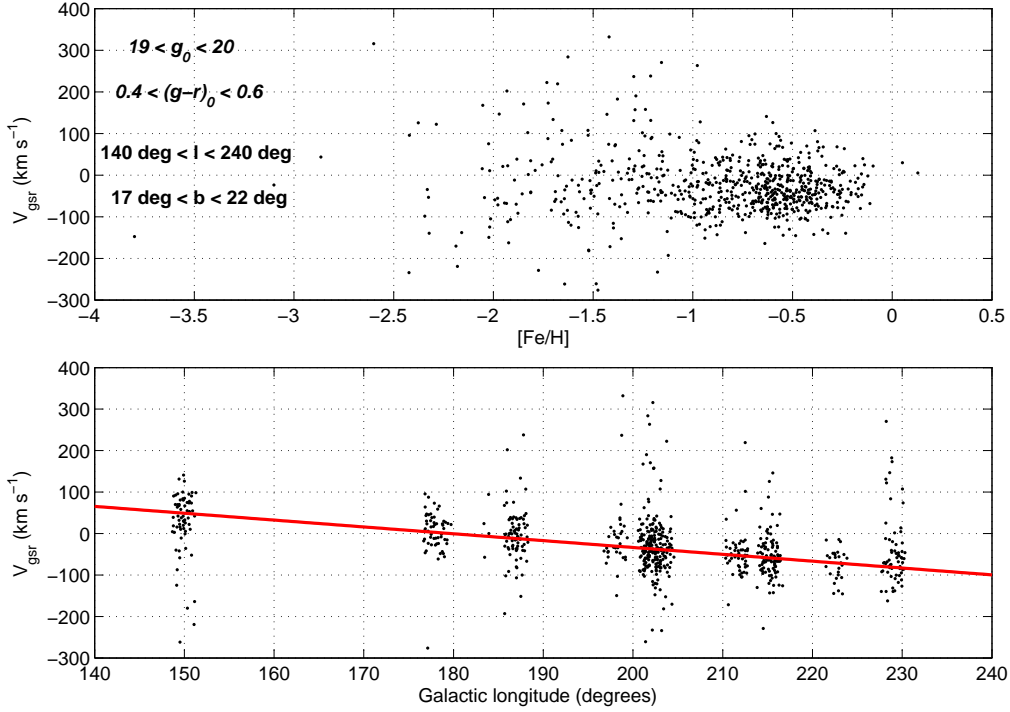}
\caption{ Line-of-sight velocities of faint stars selected to be part of the thick disk. The diagram includes only stars with spectra in SDSS DR8 which have $19<g_0<20$, $0.4<(g-r)_0<0.6$, $140\arcdeg<l<240\arcdeg$, and $17\arcdeg<b<22\arcdeg$. The top plot shows the line-of-sight, Galactic standard of rest velocity ($V_{gsr}$) as a function of metallicity. Note that low metallicity stars have large velocity dispersions, centered on zero, as one would expect for the stellar halo. The lower plot shows $V_{gsr}$ as a function of Galactic latitude. The expected $V_{gsr}$ as a function of Galactic longitude, assuming a disk rotation speed of 170 km s$^{-1}$ and a distance of 6 kpc for these stars, is shown by the red line. The velocity, velocity dispersion, and metallicity of the selected stars are in good agreement with our expectations for thick disk stars 6 kpc from the Sun and 2 kpc above the Galactic plane.}
\end{figure}

\begin{figure}
\centering
\figurenum{6}
\includegraphics[scale=0.6,angle=0]{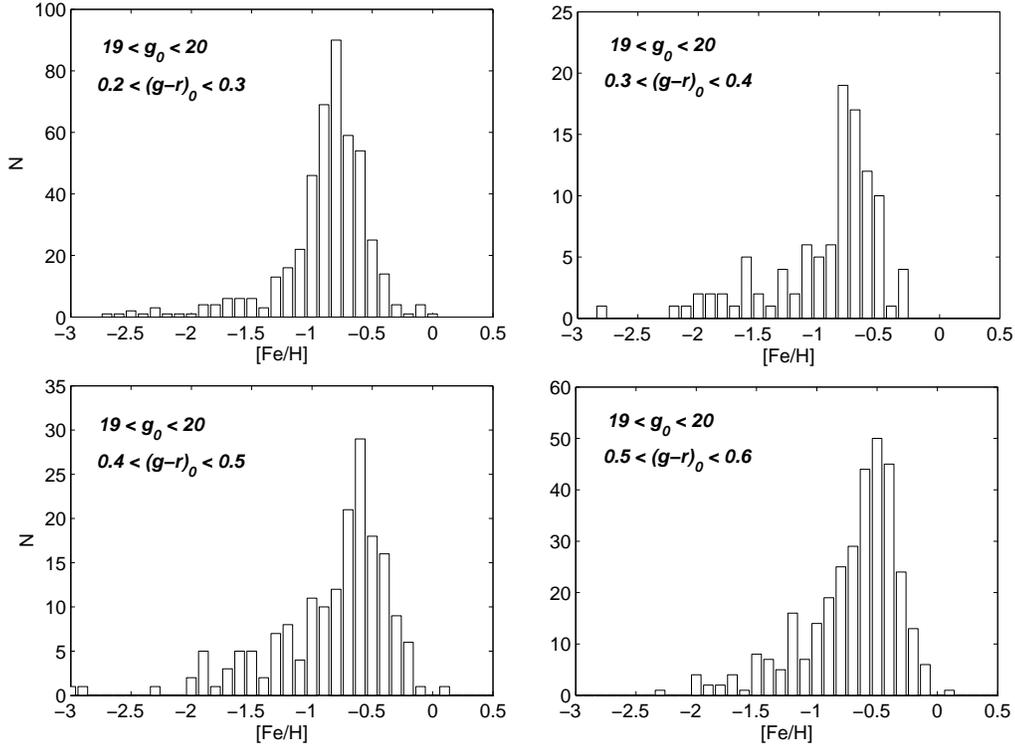}
\caption{ Metallicity vs. color for low velocity dispersion stars in the anti-center. We show the metallicity distribution of stars with SDSS DR8 spectra and which satisfy the criteria $140\arcdeg<l<240\arcdeg$, $17\arcdeg<b<22\arcdeg$, $19<g_0<20$, and which have velocities within 50 km s$^{-1}$ of the expected velocity for thick disk stars. By comparison with Figure 4, we expect the upper left panel $[0.2<(g-r)_0<0.3]$ to represent stars in the low latitude anti-center substructure. We expect the lower two panels [$0.4<(g-r)_0<0.5$ and $0.5<(g-r)_0<0.6$] to represent stars in the thick disk. And we expect stars in the upper right panel [$0.3<(g-r)_0<0.4$] to include some stars from each group; we do not believe the upper right panel indicates a gradient in the metallicity of the thick disk (see text for a detailed explanation). From this data, we conclude the thick disk has a metallicity of [Fe/H]$=-0.6$ at a distance of about 14 kpc from the Galactic center and two kpc above the Galactic plane.}
\end{figure}

\begin{figure}
\centering
\figurenum{7}
\includegraphics[scale=0.6,angle=0]{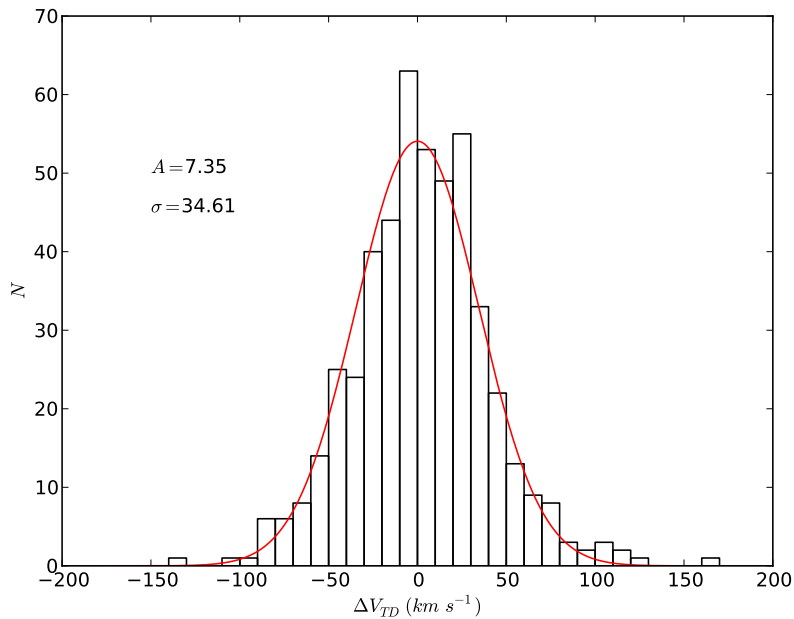}
\caption{ Thick disk stars from SDSS DR8. We selected stars with $140\arcdeg<l<240\arcdeg$, $17\arcdeg<b<22\arcdeg$, $19<g_0<20$, $0.4<(g-r)_0<0.6$, and [Fe/H]$>-1.0$. We then subtracted the expected line-of-sight velocity for a thick disk star at the $(l,b)$ of each of the observed stars, histogrammed the resulting line-of-sight velocities, and fit a Gaussian centered on zero to the peak. ``A'' is square-root of the fit amplitude. The velocity dispersion of the thick disk in the anti-center direction is $\sigma=34.6\pm 2.0$km s$^{-1}$. }
\end{figure}

\begin{figure}
\centering
\figurenum{8}
\includegraphics[scale=0.6,angle=0]{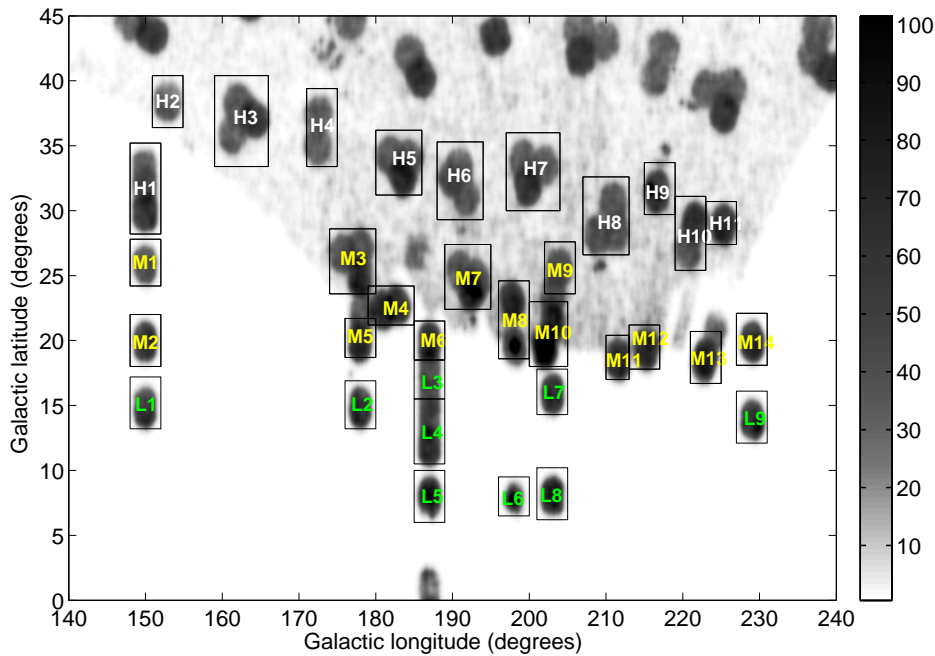}
\caption{ Selection of spectroscopic regions. We show a density map of all spectra from SDSS DR8 in the depicted region. The spectral sky coverage is highly non-uniform, and is concentrated in circular regions that were observed with SEGUE and SEGUE-2 plates. We will only use spectra in denser regions of this diagram, indicated by numbered rectangles. We will refer to the white numbered regions as the ``higher part'', the yellow numbered regions as the ``middle part'', green numbered regions as the ``lower part''.}
\end{figure}

\begin{figure}
\centering
\figurenum{9}
\includegraphics[scale=0.6,angle=0]{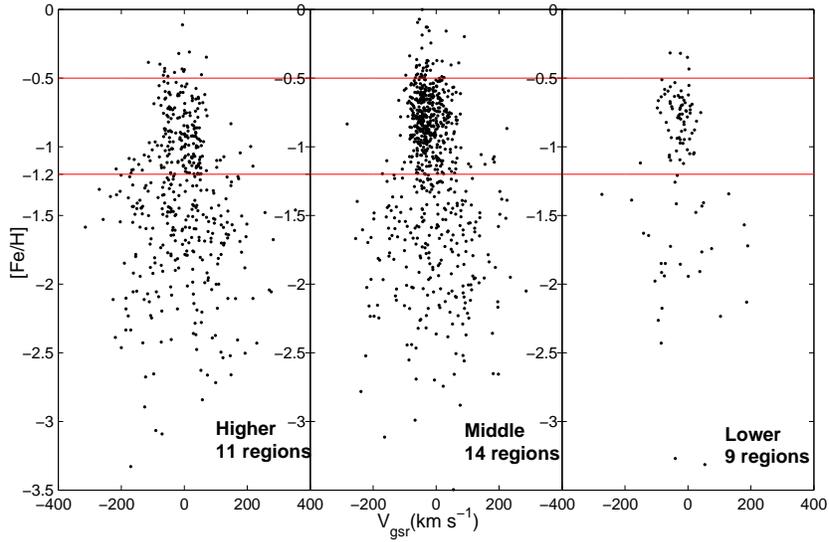}
\caption{ This figure shows the metallicity distribution of the spectra in the higher part, the middle part, and the lower part, for spectra with $0.2<(g-r)_0<0.3$ and $19<g_0<20$. The stars are clearly divided between a higher metallicity ($-1.2<$ [Fe/H] $<-0.5$, as shown by the red horizontal lines) group with low velocity dispersion and a lower metallicity group with high velocity dispersion. The high dispersion group looks like the distribution one would expect from a stellar halo population, with a mean metallicity of [Fe/H]$\sim-1.6$ and a velocity dispersion around $\sigma=120$ km s$^{-1}$. In the left panel, $55.1\%$ of the 519 stars have metallicity lower than $-1.2$. In the middle panel, $30.0\%$ of 677 stars have metallicity lower than $-1.2$, and in the right panel, $29.8\%$ of 104 stars have metallicity lower than $-1.2$.}
\end{figure}

\begin{figure}
\centering
\figurenum{10}
\includegraphics[scale=0.6,angle=0]{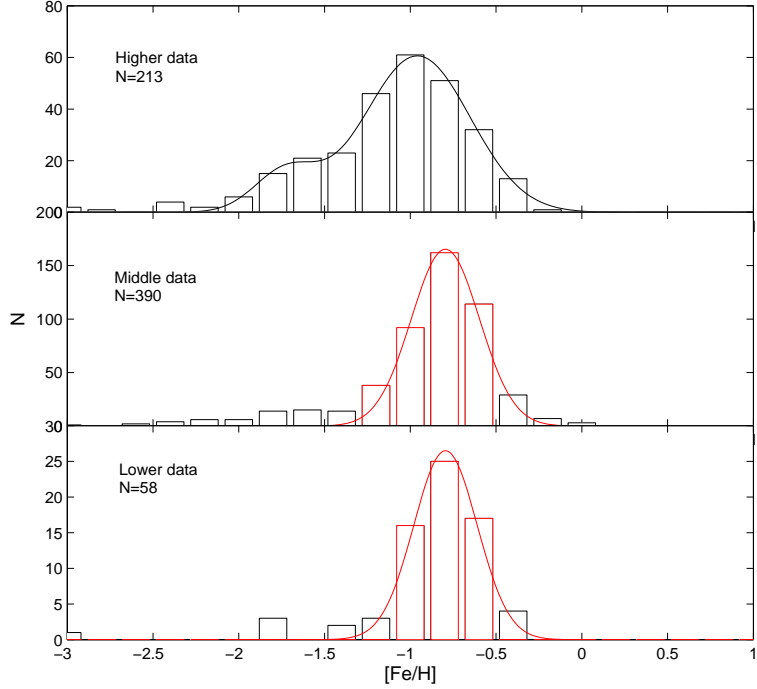}
\caption{ Substructure metallicity distribution. We separated the stars with $19<g_0<20$ and $0.2<(g-r)_0<0.3$ by membership in the higher, middle, and lower regions (higher, middle, and lower panels, respectively), and used only spectra with $V_{gsr}$ within 50 km s$^{-1}$ of the expected thick disk velocity in the given direction (see red curves in Figure 11). Note that the spectra in the lower and middle regions have almost an identical metallicity distribution of [Fe/H]$=-0.80\pm 0.01$ and [Fe/H]$=-0.80\pm 0.02$. We fit two Gaussians to the higher region data. We fit a single Gaussian to the data in the red bins in the middle and lower region data. The higher plates have a lower metallicity, of [Fe/H]$=-0.96\pm 0.03$. The higher plates also have a slightly wider range of metallicities.}
\end{figure}

\begin{figure}
\centering
\figurenum{11}
\includegraphics[scale=0.4,angle=0]{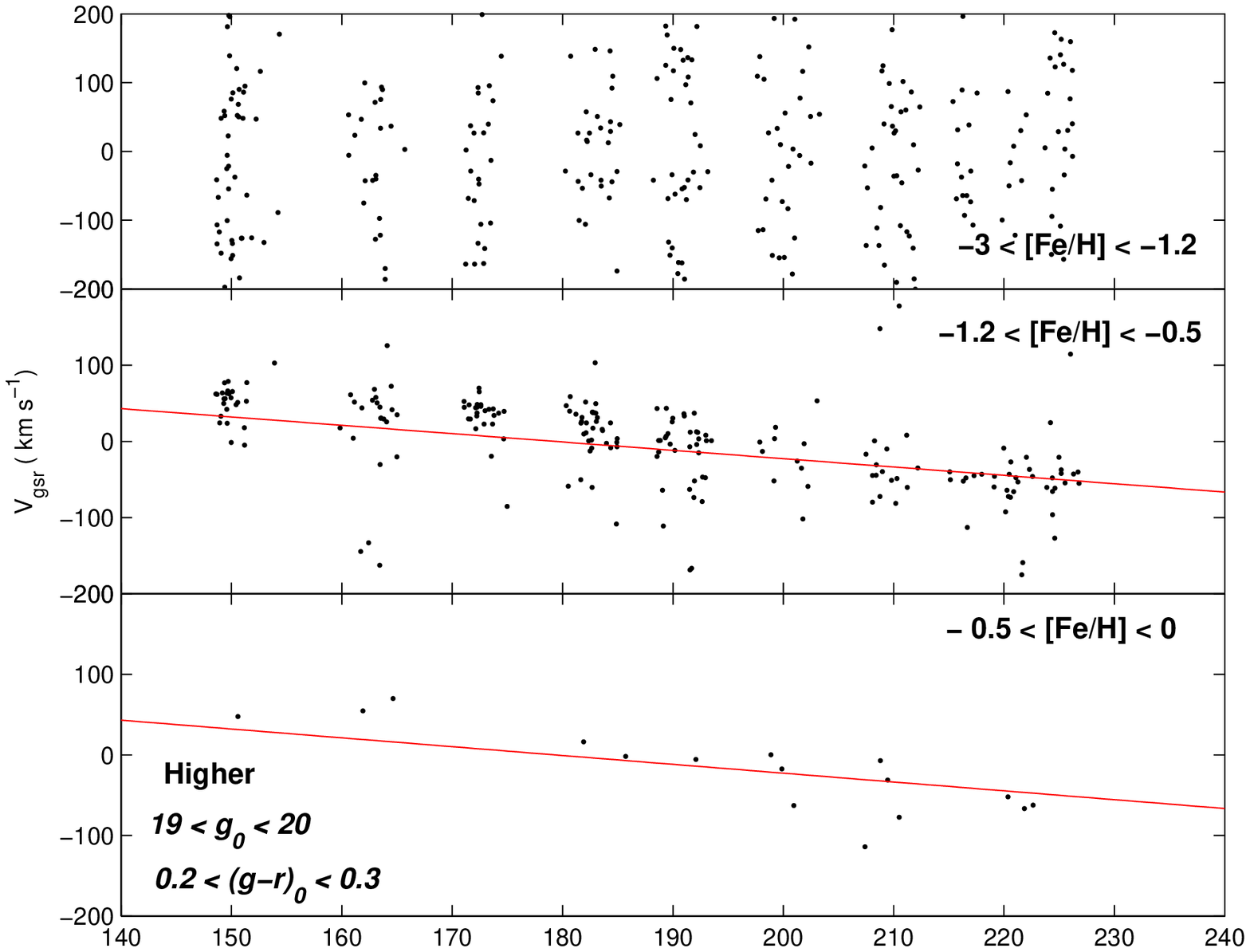}
\includegraphics[scale=0.4,angle=0]{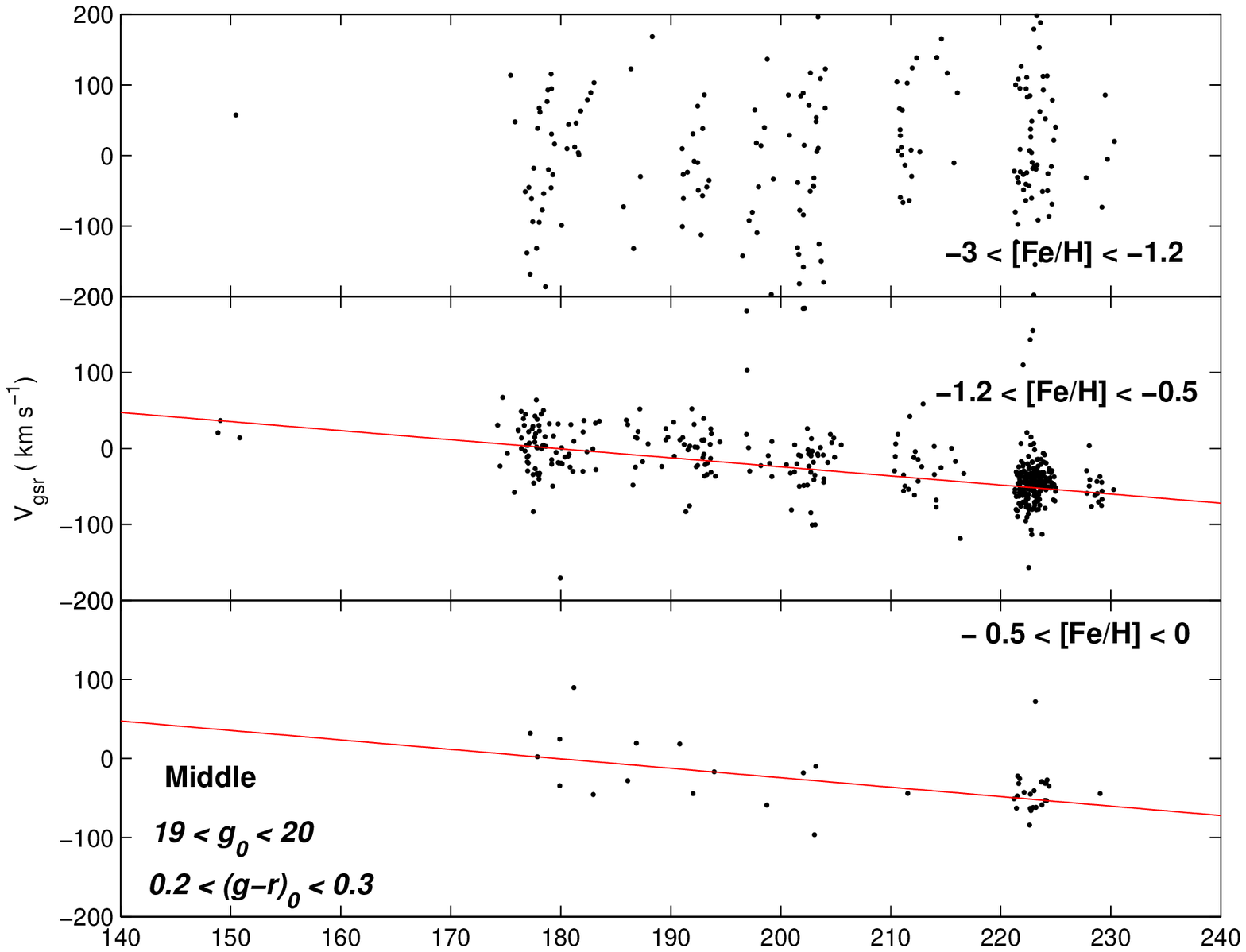}
\includegraphics[scale=0.4,angle=0]{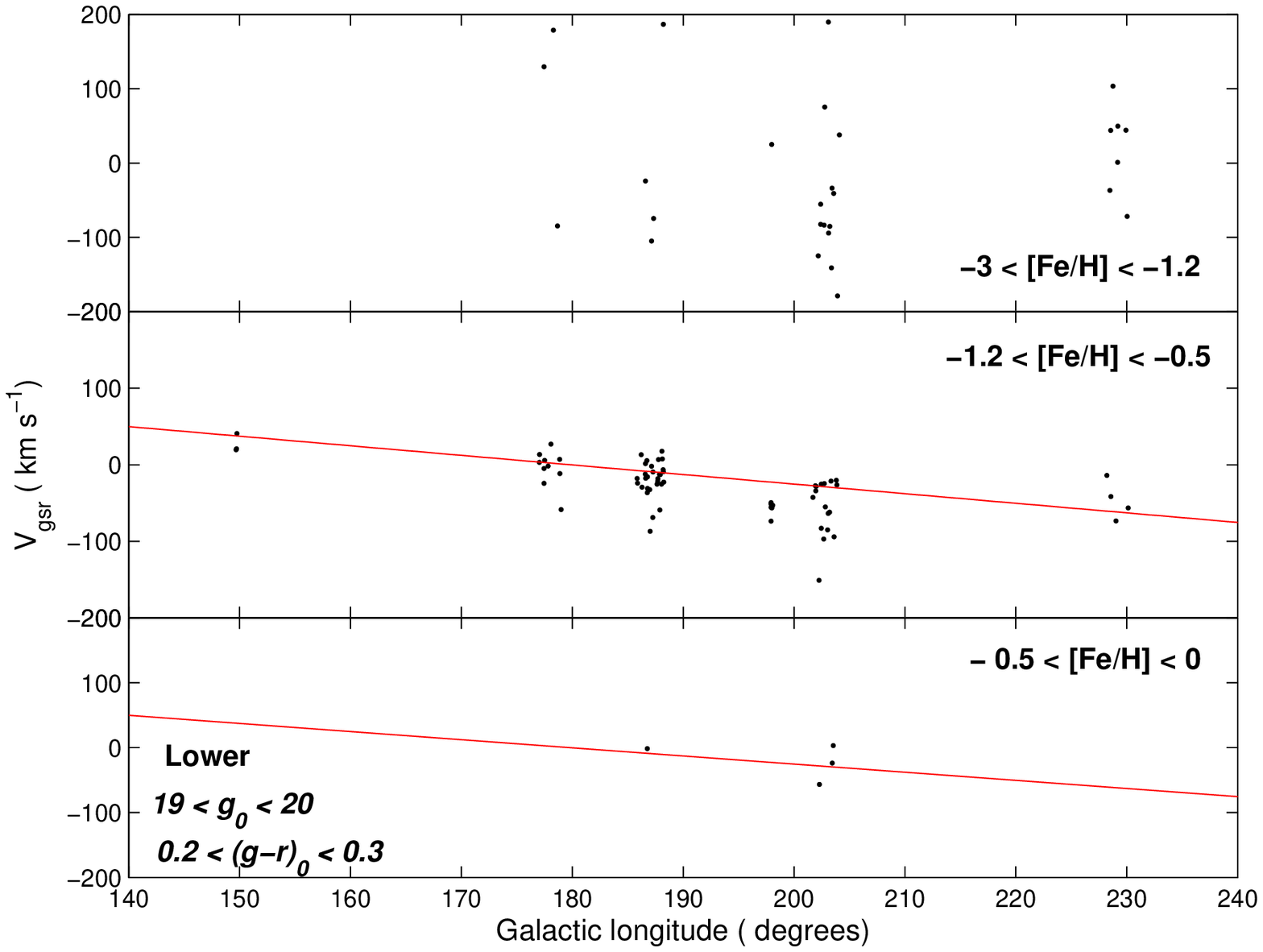}

\caption{Velocity as a function of Galactic longitude for higher, middle, and lower spectra, separated by metallicity. The red line shows the expected velocity of the thick disk at 10 kpc from the Sun. Note that in general the low metallicity stars match expectations for the stellar spheroid --- a broad velocity distribution centered at $V_{gsr}$=0. The intermediate metallicity higher part data has higher line-of-sight velocities than one would expect for a thick disk, regardless of the distance of the stars; there stars are from the ACS. The lower part intermediate metallicity stars have the mean velocities expected for the thick disk, but a narrower velocity dispersion. The lower part also has velocities at $l=203\arcdeg$ that are lower than one would expect for the thick disk; these stars are likely part of the leading tidal tail of the Sagittarius dwarf tidal stream. The middle part has a larger velocity dispersion than either of the other two sets of data; these stars may be a combination of ACS and Monoceros Ring stars. There are very few high metallicity stars, but the generally follow the trends of the intermediate metallicity stars. The lower part intermediate metallicity stars have the mean velocities expected for the thick disk, but a narrower velocity dispersion; these are stars from the Monoceros Ring.}
\end{figure}

\begin{figure}
\centering
\figurenum{12}
\includegraphics[scale=0.5,angle=0]{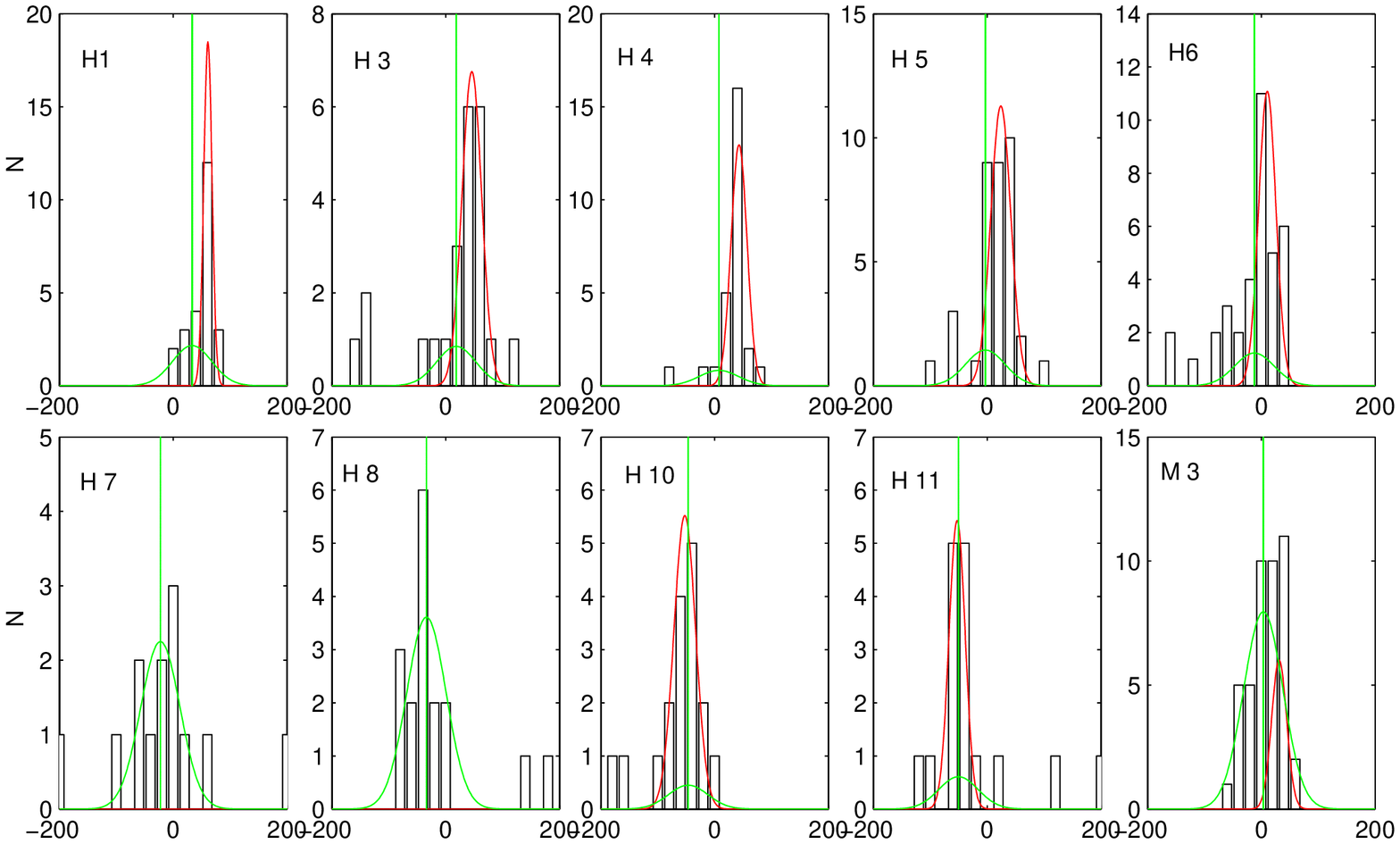}
\includegraphics[scale=0.5,angle=0]{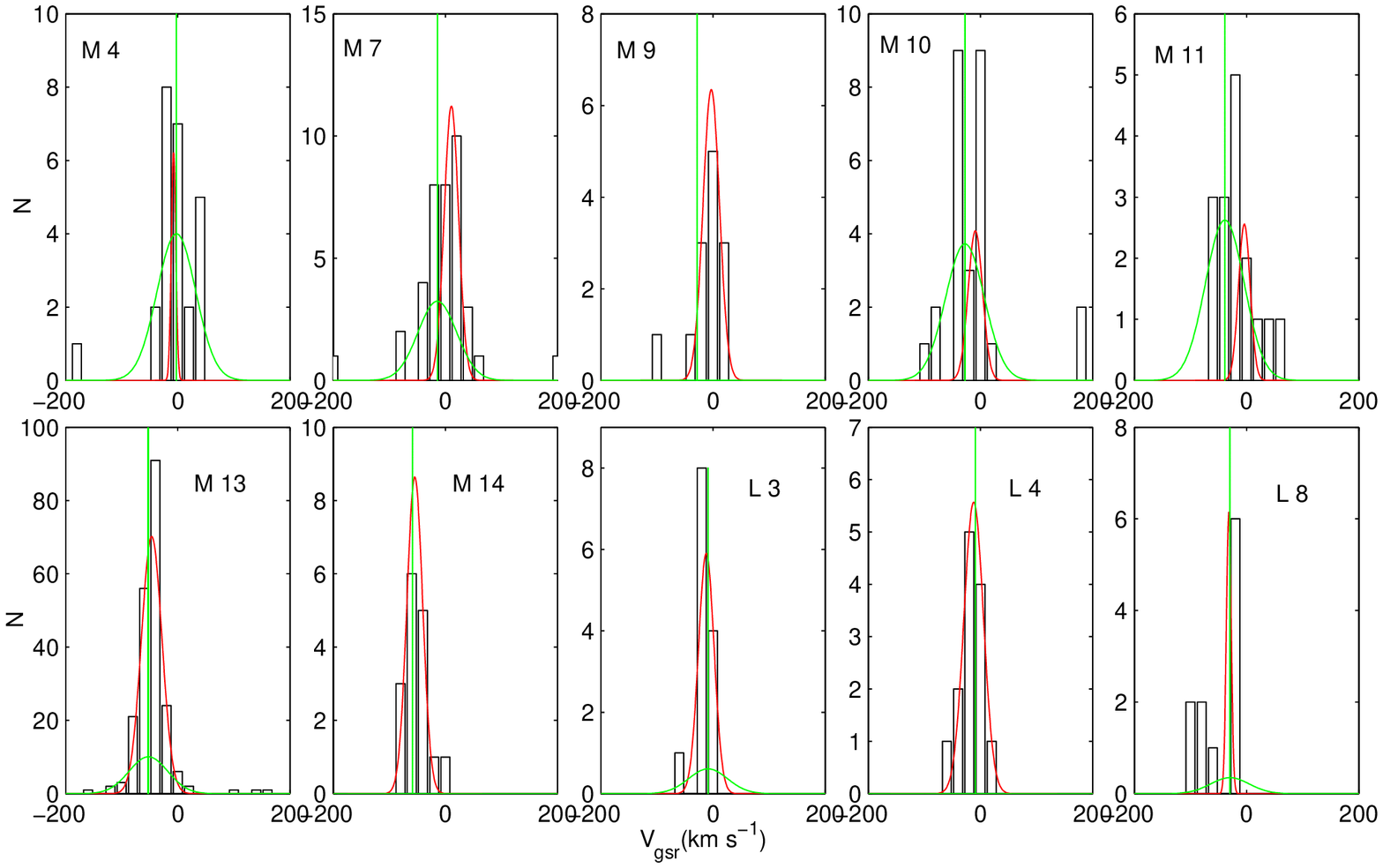}
\caption{Velocity distribution of substructure by region of sky. All of the stars selected in each region have $19<g_0<20$, $0.2<(g-r)_0<0.3$, and $-1.2<$[Fe/H]$<-0.5$. The velocity distribution in each panel was fit with two Gaussians : one with the velocity and dispersion of the thick disk (green line), and the other with a floating mean velocity and dispersion (red line). From this figure and the fit parameters Table 2, one concludes that H1, H3, H4, H5, H6, M7, and M9 have a narrow, shifted peak (pink in Figure 13). A significant fraction of the stars in M3, M10, and M11 are well fit to our expectations for the thick disk, but they are better fit if a fraction of the stars are assigned to a narrow, shifted peak (blue in Figure 13). H7,H8, M4 are consistent with just thick disk (black in Figure 13). Most of the stars in H10, H11, M13, M14, L3, L4 are consistent with a peak at the same velocity as the TD, but have a significantly narrower velocity dispersion (green in Figure 13). Most of the panels that show a narrow velocity peak also have a few stars that are fit by a broader thick disk component (green line).}
\end{figure}

\begin{figure}
\centering
\figurenum{13}
\includegraphics[scale=0.6,angle=0]{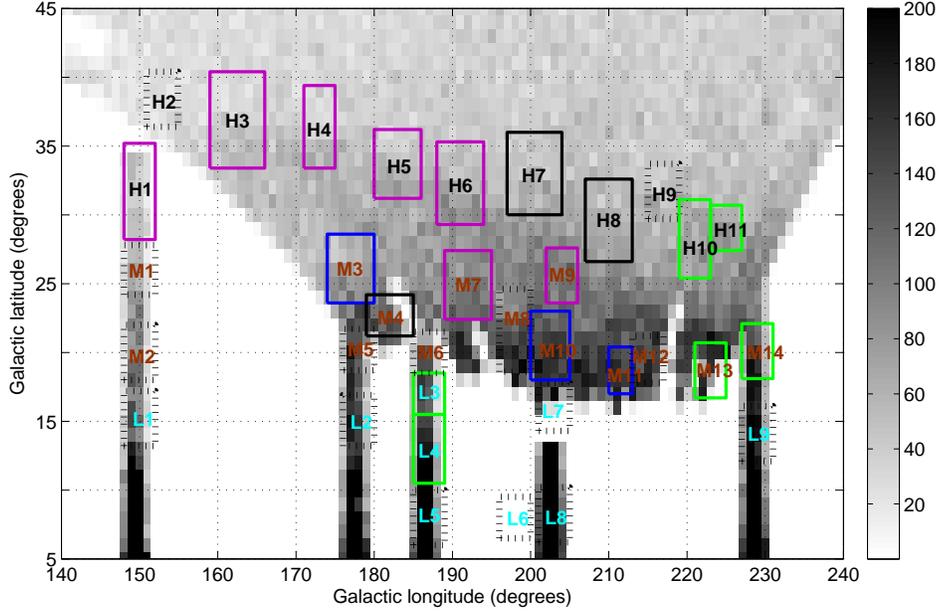}
\caption{Sky area separated by line-of-sight velocity. Regions are color coded by how well the velocities of the blue stars match the velocities expected for the thick disk. The colors indicate: pink -- narrow peak velocity, different from thick disk; green --- narrow peak velocity, similar to thick disk; black --- velocity peak consistent with thick disk; blue --- most of the stars consistent with thick disk, but better fit with an extra, shifted narrow peak. Notice that the low latitude and EBS (higher plates 10 and 11) regions of the sky have narrow peaks with thick disk-like mean velocities. The higher latitude portions of the diagram, including both the areas tracking the sharp edge of the stellar over-density and a region of the sky at lower latitude than the published position of the ACS (middle plate 7) have velocities that are higher than expected for the thick disk (and inconsistent with any circularly symmetric, long-lived stellar substructure). Where there is no over-density of blue turnoff stars (higher plates 7 and 8), we have velocity dispersions and velocities that are plausibly consistent with the thick disk. At low latitude, the data is consistent with thick disk rotation, but with a significantly narrower velocity dispersion.}
\end{figure}

\begin{figure}
\centering
\figurenum{14}
\includegraphics[scale=0.6,angle=0]{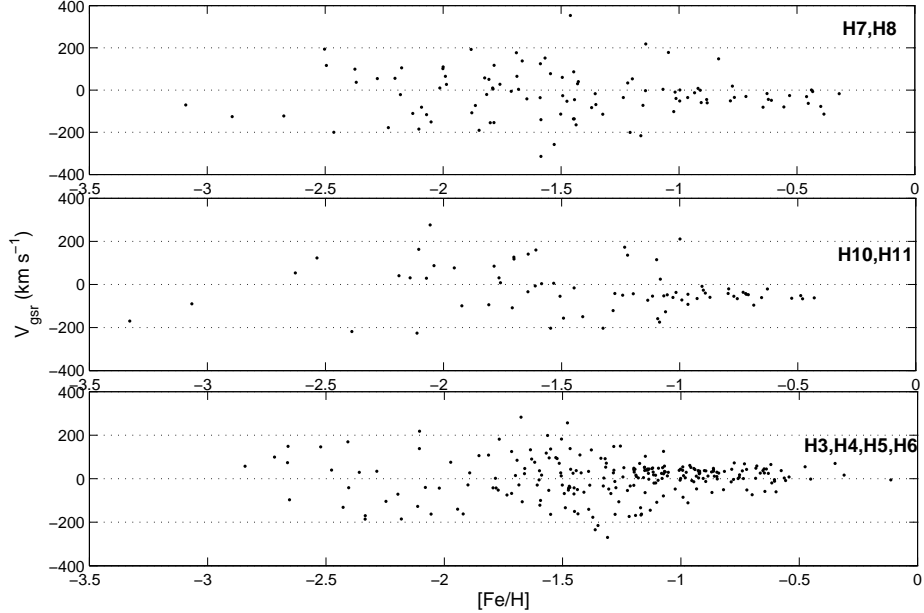}
\caption{$V_{gsr}$ vs. metallicity for stars in separate high latitude structures. Stars were selected with $19<g_0<20$ and $0.2<(g-r)_0<0.3$. Panels include high latitude regions with no apparent anti-center substructure (plates H7, H8, top panel), regions in the EBS (plates H10, H11, middle panel), and regions containing the ACS (plates H3, H4, H5, H6, lower panel). In all three panels, there are a significant number of stars that belong to the spheroid, with low metallicity and a large velocity dispersion. All three panels have a narrower velocity dispersion component at [Fe/H]$>-1.2$. Notice that in the top panel, there are as many stars in the narrow component with [Fe/H]$\sim-0.5$ as there are at [Fe/H]$\sim-0.9$, and hardly any stars in the narrow velocity dispersion component that are more metal-poor than [Fe/H]$=-1.0$. In the ACS (lower panel), most of the narrow component stars are near [Fe/H]$=-1.0$, and there are hardly any stars near [Fe/H]$=-0.5$. Also, most of the narrow velocity dispersion stars have positive velocities, and the data is consistent with an additional slightly broader velocity dispersions component centered around $V_{gsr}=0$ km s$^{-1}$, as we expect for the thick disk. The EBS stars (middle panel) have a metallicity closer to [Fe/H]$=-0.8$, and the velocity dispersion is narrower than the region without the EBS. The mean line-of-sight velocity of the EBS stars is the same as expected for the thick disk, so it is difficult to tell if there are a few stars with a slightly larger dispersion in the mix.}
\end{figure}

\begin{figure}
\centering
\figurenum{15}
\includegraphics[scale=0.7,angle=0]{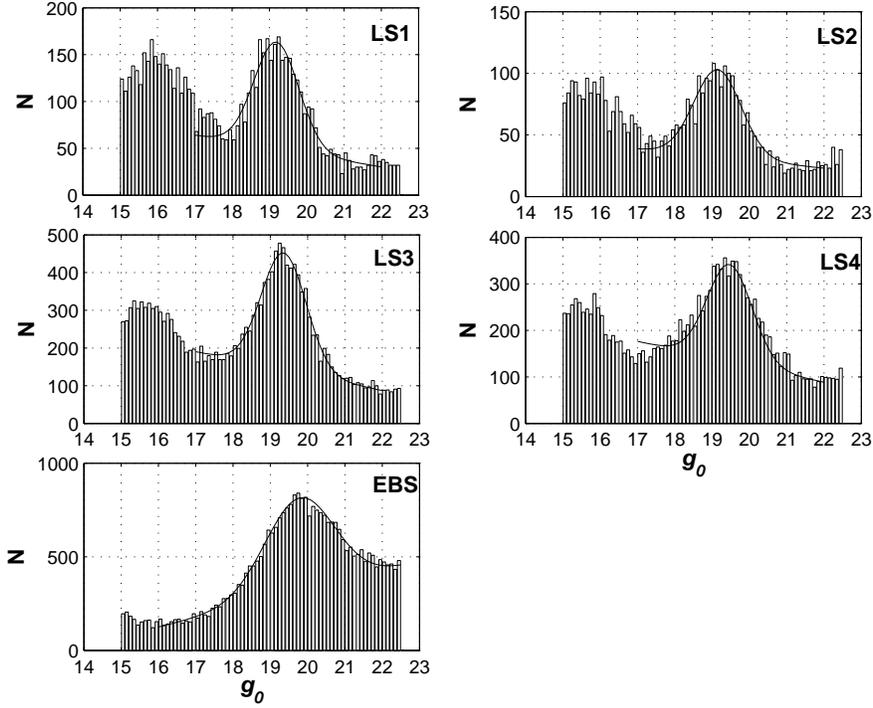}
\caption{Magnitude distribution in several lines of sight over the Monoceros Ring and EBS. Stars with $0.2<(g-r)_0<0.3$ were selected in four regions of sky along the Anti-Center Stream, chosen to include low latitude SEGUE stripes. The areas of the sky selected are: LS1 --- $170\arcdeg<l<180\arcdeg$, $15\arcdeg< b<18\arcdeg$; LS2 --- $180\arcdeg<l<190\arcdeg$, $15\arcdeg< b<18\arcdeg$; LS3 --- $200\arcdeg<l<210\arcdeg$, $15\arcdeg< b<18\arcdeg$; LS4 --- $220\arcdeg<l<230\arcdeg$, $15\arcdeg< b<18\arcdeg$; and EBS - $220\arcdeg<l<230\arcdeg$, $(-0.15l+62.1\arcdeg)<b<(-0.15l+64.6\arcdeg)$. These regions are outlined in blue and red in Figure 16. The center of the peak in the magnitude distribution is fit with a line plus a Gaussian. The peak apparent magnitude is $g_0=$19.19, 19.15, 19.36, 19.47, and 19.88, in regions LS1, LS2, LS3, LS4, and EBS, respectively (though see Section 6.3 and Figures~22 and 23 for caveats about the EBS peak magnitude).}
\end{figure}

\clearpage
\begin{figure}
\centering
\figurenum{16}
\includegraphics[scale=0.6,angle=0]{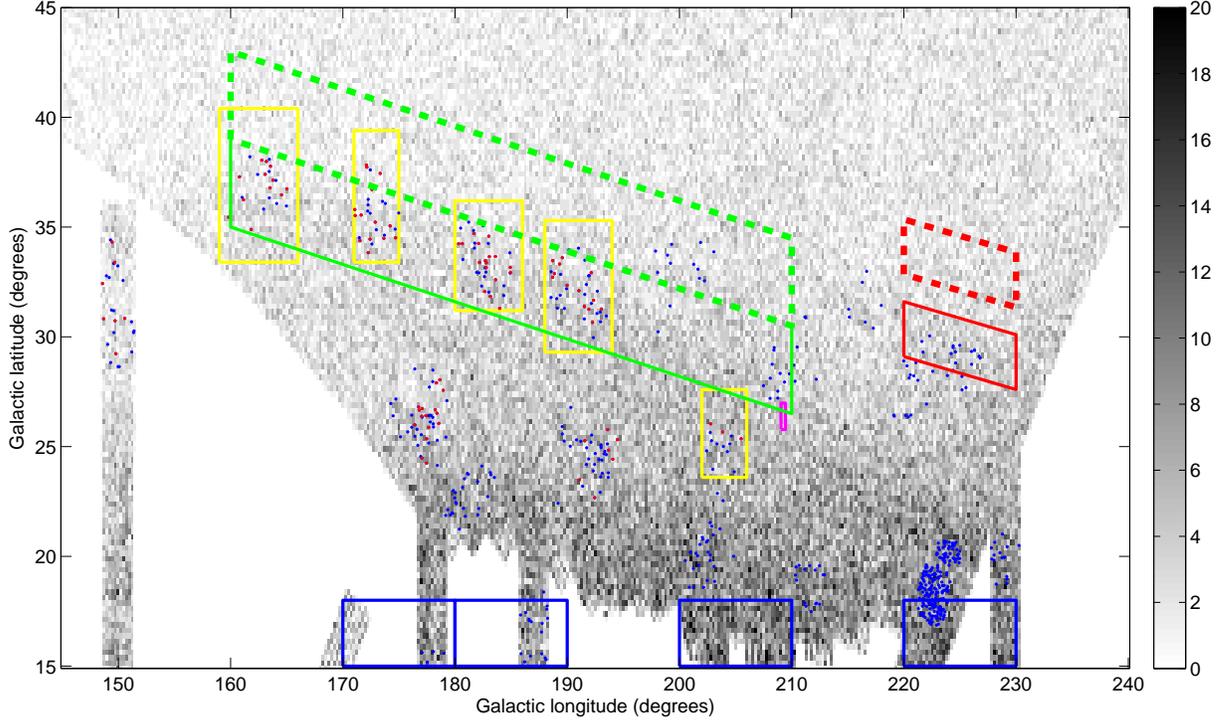}
\caption{Positions of stars in the Anti-Center Stream. The background shows the density of blue turnoff stars with $19<g_0<20$, $0.2<(g-r)_0<0.3$, and $(u-g)_0>0.4$. The bin size is 0.25 degrees in Galactic latitude and longitude (so the actual sky area is smaller for larger Galactic latitudes). The red points show the positions of spectra with $-1.2<$[Fe/H]$<-0.9$, $19<g_0<20$, $0.2<(g-r)_0<0.3$, and velocities within 2 sigma of the narrow peak in regions H1, H3, H4, H5, H6, M3, M7, and M9, as shown in Figure 12 and tabulated in Table 2. The green outlines show the areas of the sky used to determine the distances to the ACS (solid line), and the reference region (dashed line) used for comparison. The yellow squares show regions H3, H4, H5, H6, and M9, which were used to determine the line-of-sight $V_{gsr}$ of the ACS. The red and blue outlines show the sky regions used to determine the distance to the EBS and Monoceros ring, respectively. The pink square near $(l,b)=(209\arcdeg,26\arcdeg)$ shows the position of the data used by \citet{2010ApJ...725.2290C} to fit the proper motion of the ACS.}
\end{figure}

\begin{figure}
\centering
\figurenum{17}
\includegraphics[scale=0.5,angle=0]{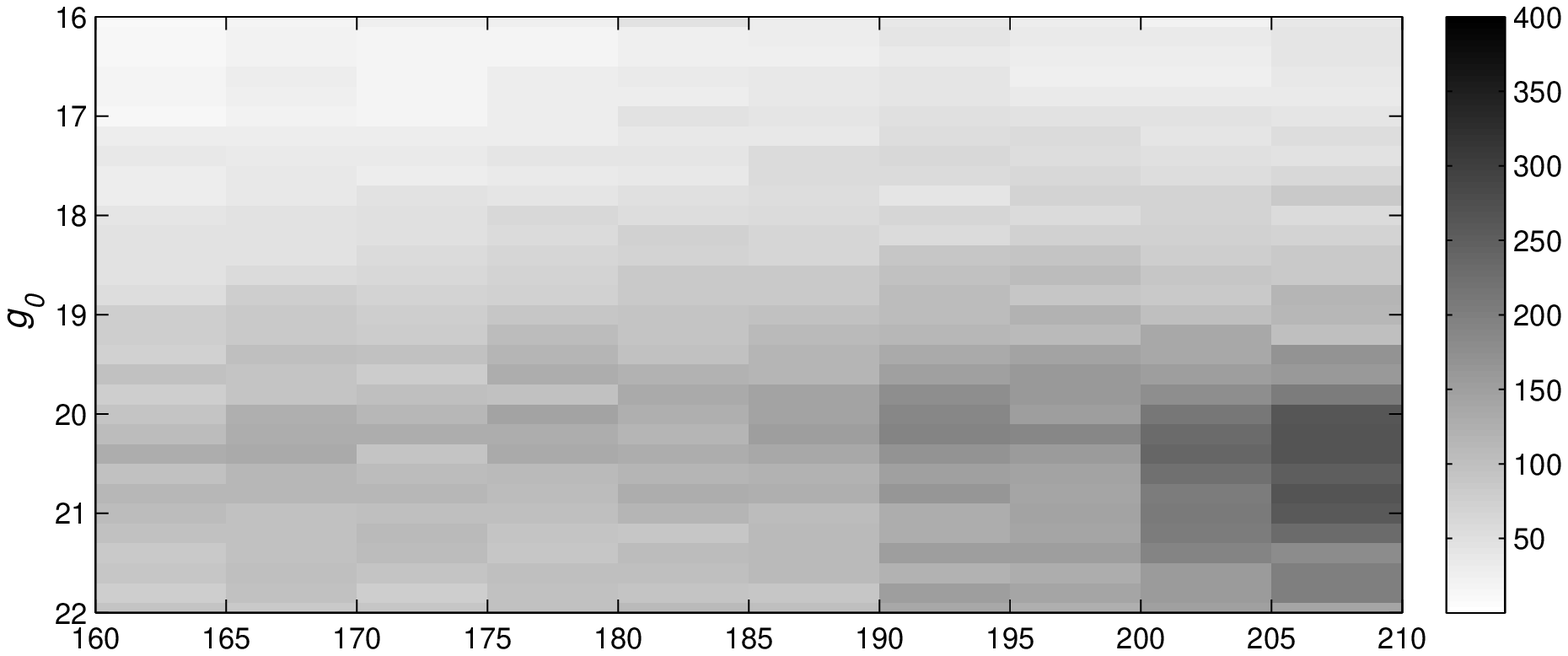}
\includegraphics[scale=0.5,angle=0]{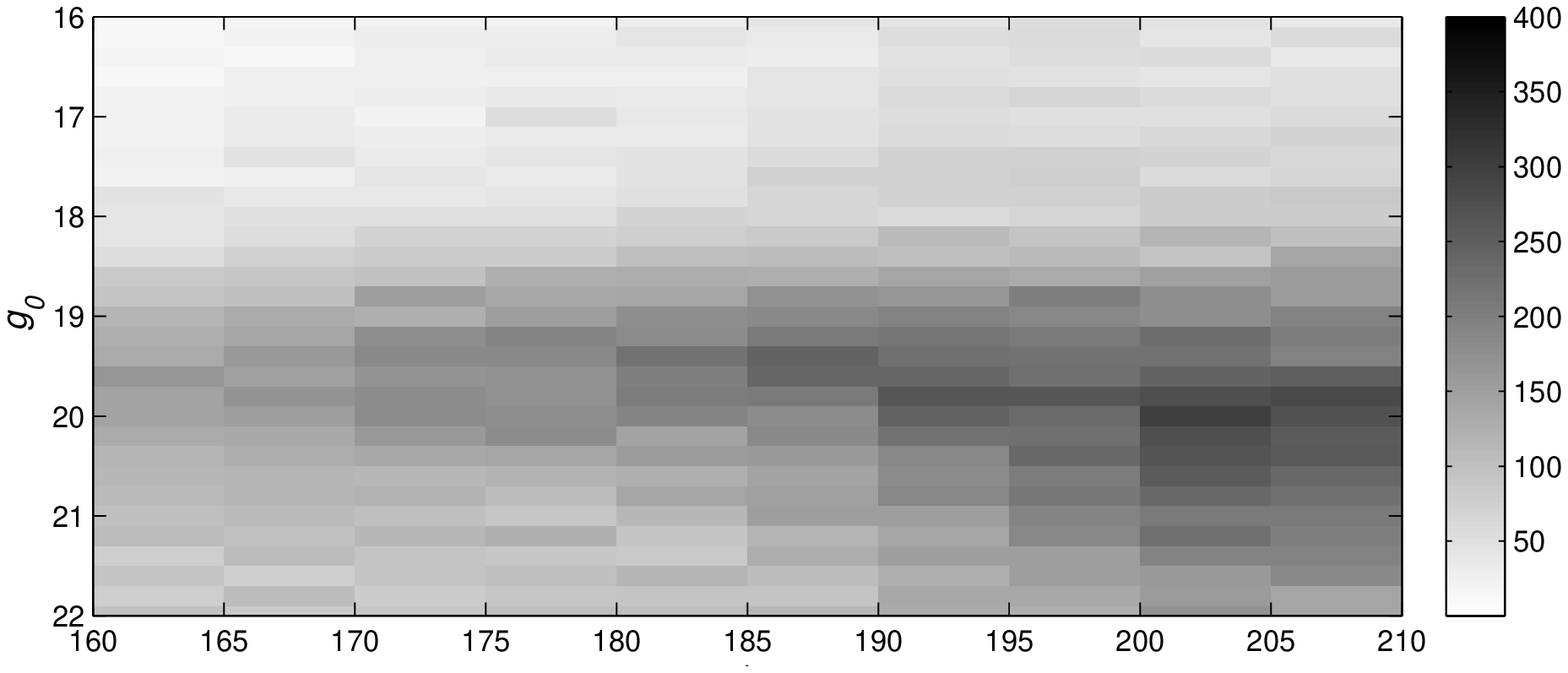}
\includegraphics[scale=0.5,angle=0]{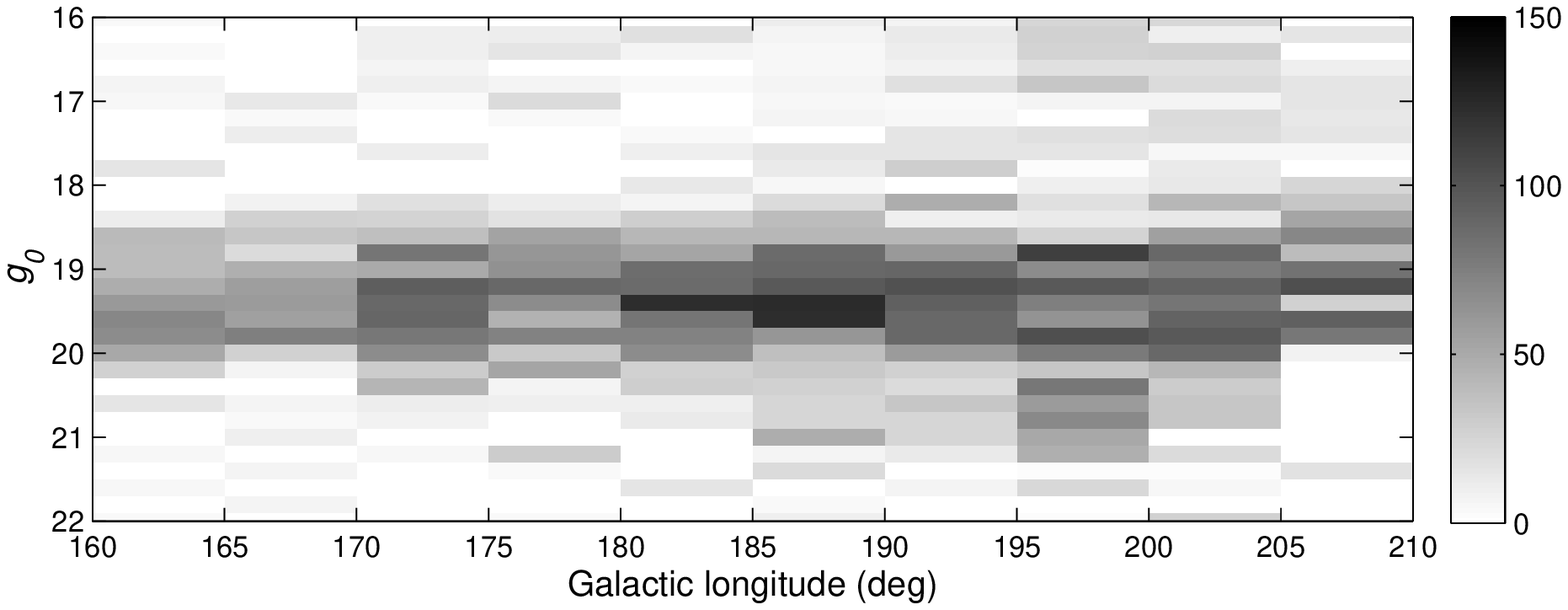}
\caption{Determining the distance to the ACS. All stars used in this figure were selected with $0.2<(g-r)_0<0.3$ and $(u-g)_0>0.4$. The top panel shows the apparent $g_0$ magnitude distribution for stars with $-0.17l+66.2\arcdeg<b<-0.17l+70.2\arcdeg$. These stars are above the ACS (see Figure 16). Note the extra density of Sagittarius leading tidal tail stars at $(l,b)=(205\arcdeg,20.5\arcdeg$). The middle panel shows stars on the top edge of the ACS, with $-0.17l+62.2\arcdeg<b<-0.17l+66.2\arcdeg$, The magnitude distribution at higher Galactic latitude is contaminated with Sagittarius stream stars. To get a purer sample of ACS stars, we subtracted the upper panel from the middle panel to produce the lower panel. The subtraction of the Sagittarius stars is not perfect, because the Galactic longitude of the stream increases slightly with Galactic latitude. However, one can see that the distance to the ACS is approximately constant with Galactic longitude for $168\arcdeg<l<210\arcdeg$. }
\end{figure}

\clearpage
\begin{figure}
\centering
\figurenum{18}
\includegraphics[scale=0.6,angle=0]{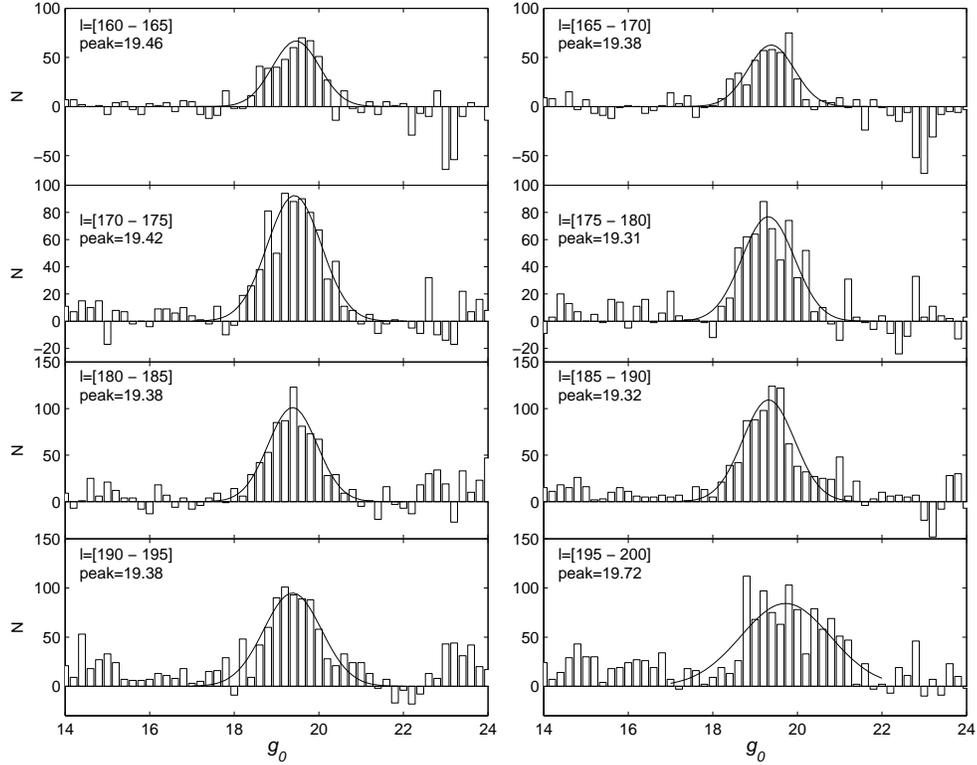}
\caption{Fit to the distance to the ACS. We fit a Gaussian to the subtracted apparent magnitude distribution in the lower panel of Figure 17. For the seven panels with $160\arcdeg<l<195\arcdeg$, the apparent magnitude of the peak and the width of the Gaussian are fairly constant. At $195\arcdeg<l<200\arcdeg$, we suspect the data are contaminated by slightly fainter Sagittarius stream stars, which broaden the magnitude distribution and cause the peak to appear fainter.}
\end{figure}

\clearpage
\begin{figure}
\centering
\figurenum{19}
\includegraphics[scale=0.6,angle=0]{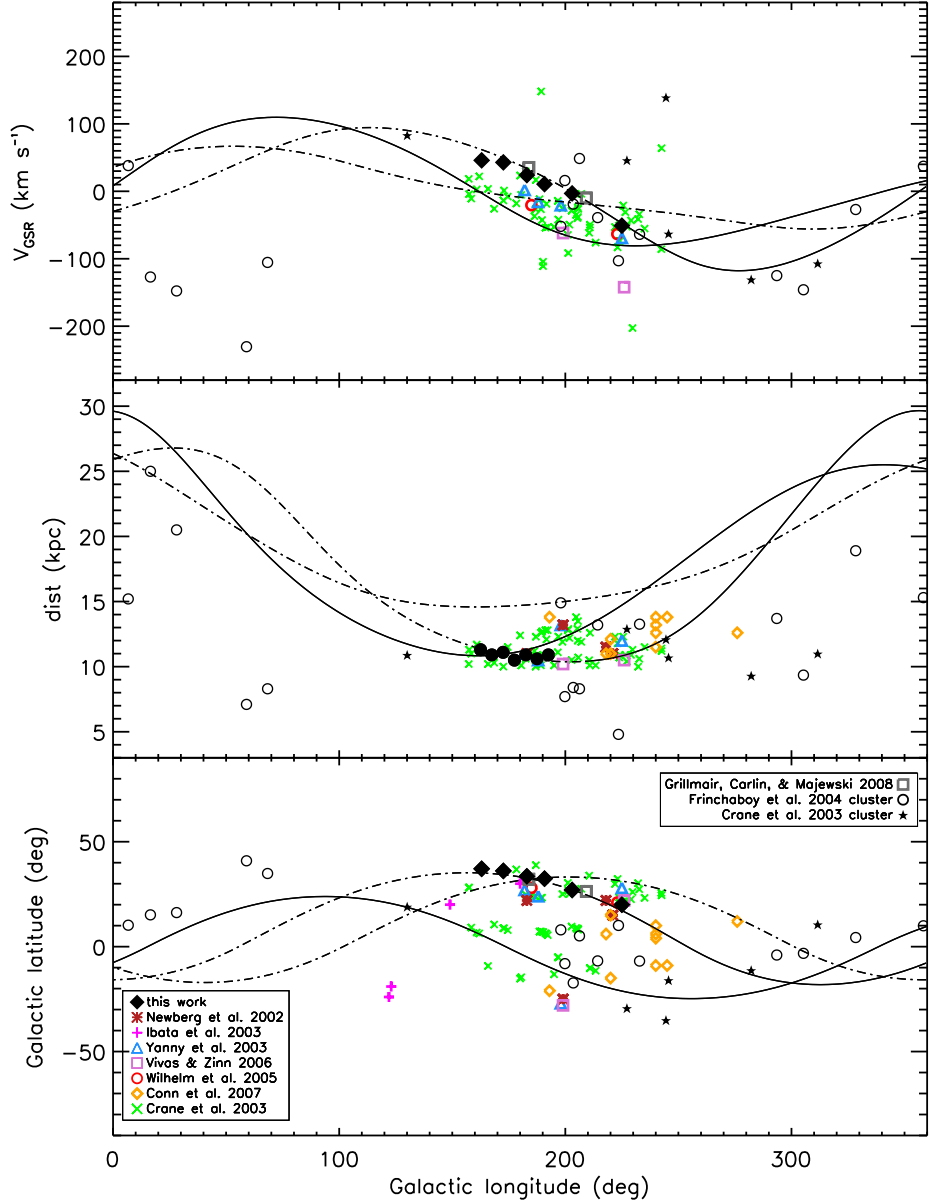}
\caption{The orbit of the Anti-Center Stream. The measurements from this paper are shown as filled diamonds in velocity and position, and filled circles in distance. Previously claimed detections of anti-center substructure, which we now know includes the ACS, EBS, and Monoceros, are given as various colored symbols. Open circles and filled black stars are open and globular clusters previously claimed to be associated with the anti-center substructures, and the two open gray squares are fields where \citet{2008ApJ...689L.117G} and subsequently \citet{2010ApJ...725.2290C} studied ACS kinematics. The orbit fit by \citet{2008ApJ...689L.117G} matches our data very well. The forward integration is the solid line, and the backward integration is the dot-dashed line.}
\end{figure}

\clearpage
\begin{figure}
\centering
\figurenum{20}
\includegraphics[scale=0.65,angle=0]{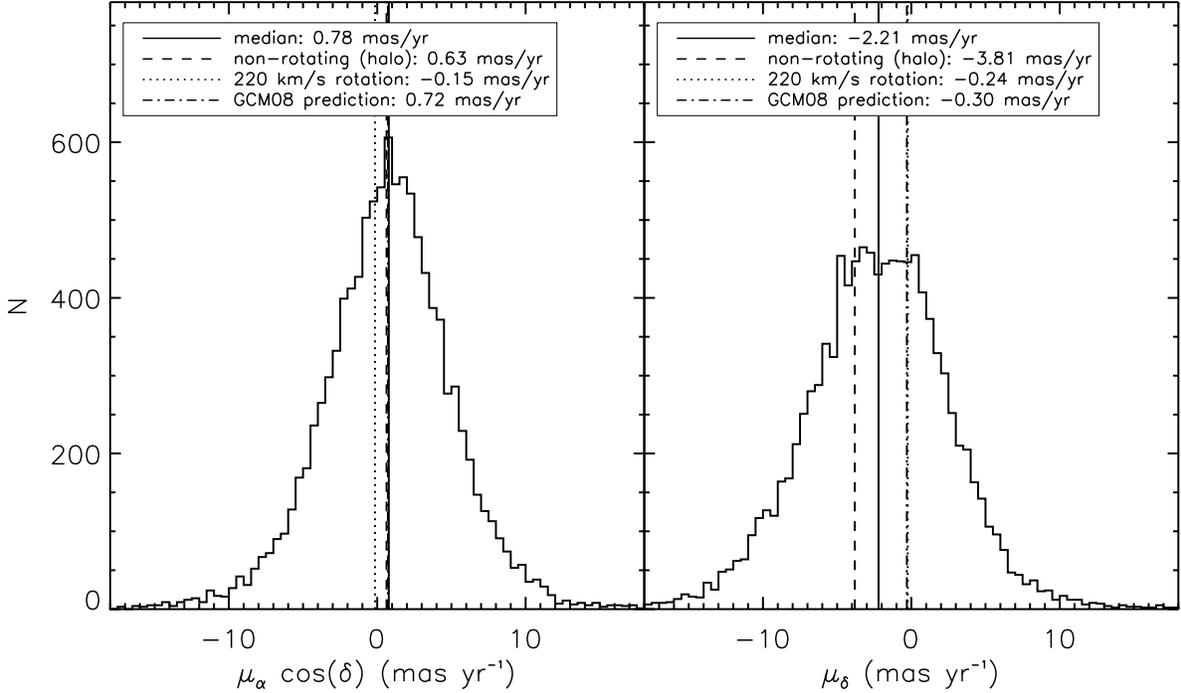}
\caption{Proper motion measurements of stars in the ACS. The stars were selected from SDSS DR7 with $160\arcdeg<l<190\arcdeg$, $29\arcdeg<b<38\arcdeg$,$19<g_0<20$,$0.2<(g-r)_0<0.3$, and $(u-g)_0>0.4$. We expect that half of the stars selected with these criteria will be ACS stars, and the other half will be halo stars. The median proper motion is show by the solid line. For reference, the expected proper motions for the non-rotating halo(dashed line), disk-like rotation of 220 km s$^{-1}$ (dotted line), and the \citet{2008ApJ...689L.117G} orbit fit (dot-dashed lines) are also shown. From the $\mu_{\delta}$ measurement, we see that a counter-rotating orbit (which would produce proper motions to the left of the dashed line) is ruled out. The broad peak is quite consistent with half of the stars peaked at the proper motions of the halo, and the other half peaked at the expected proper motion of the ACS.}
\end{figure}

\clearpage
\begin{figure}
\centering
\figurenum{21}
\includegraphics[scale=0.5,angle=0]{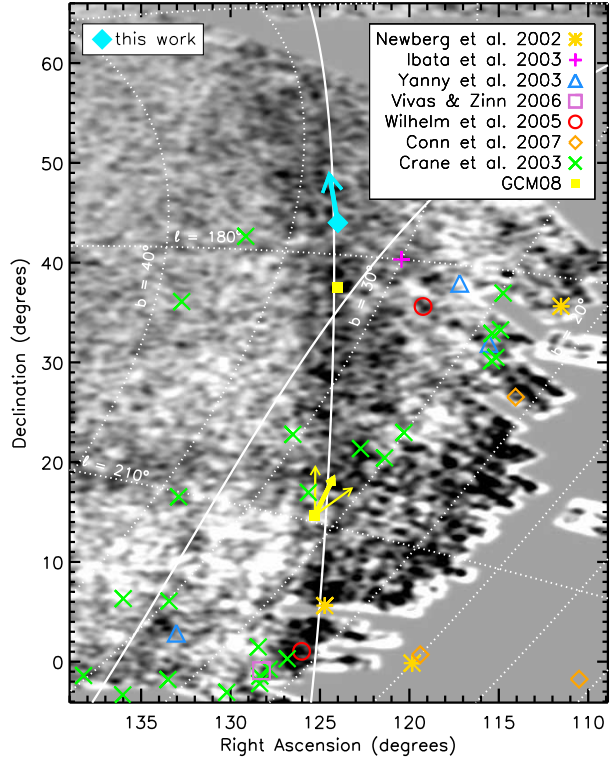}
\caption{Map of SDSS DR7 filtered star counts in the anti-center region, similar to Figure 14 of \citet{2010ApJ...725.2290C}. The Anti-center Stream is the roughly vertical swath of stars in the middle of the plot. The best-fitting orbit from GCM08 is overlaid as a solid line; this orbit was constrained to fit 30 positions along the stream and radial velocities in the two fields shown as filled (yellow) squares. The
(yellow) square at $(\alpha, \delta)\approx(125.3\arcdeg,14.7\arcdeg$) represents SA 76, the field studied in Carlin et al. (2010); the vector and its flanking 3$\sigma$ error vectors attached to this
square correspond to our revised estimate of the proper motion of substructure stars in that field (see the text for details). The large (cyan) diamond at $(\alpha, \delta)\approx(124.0\arcdeg,44.0\arcdeg$) is the mean position of the SDSS DR7 proper motion sample used in the analysis of
Fig.~20. Attached to this symbol is a vector illustrating the direction of motion implied by the median proper motion and radial velocity of stars between $19 < g_0 < 20, 0.2 < (g-r)_0 < 0.3,
160\arcdeg < l < 190\arcdeg, 29\arcdeg < b < 38\arcdeg$. This direction of motion is consistent with the (prograde) GCM08 orbit. Colored symbols show the rather scattered positions of previously
claimed detections of ``Monoceros" debris in this area of sky.}
\end{figure}

\clearpage
\begin{figure}
\centering
\figurenum{22}
\includegraphics[scale=0.7,angle=0]{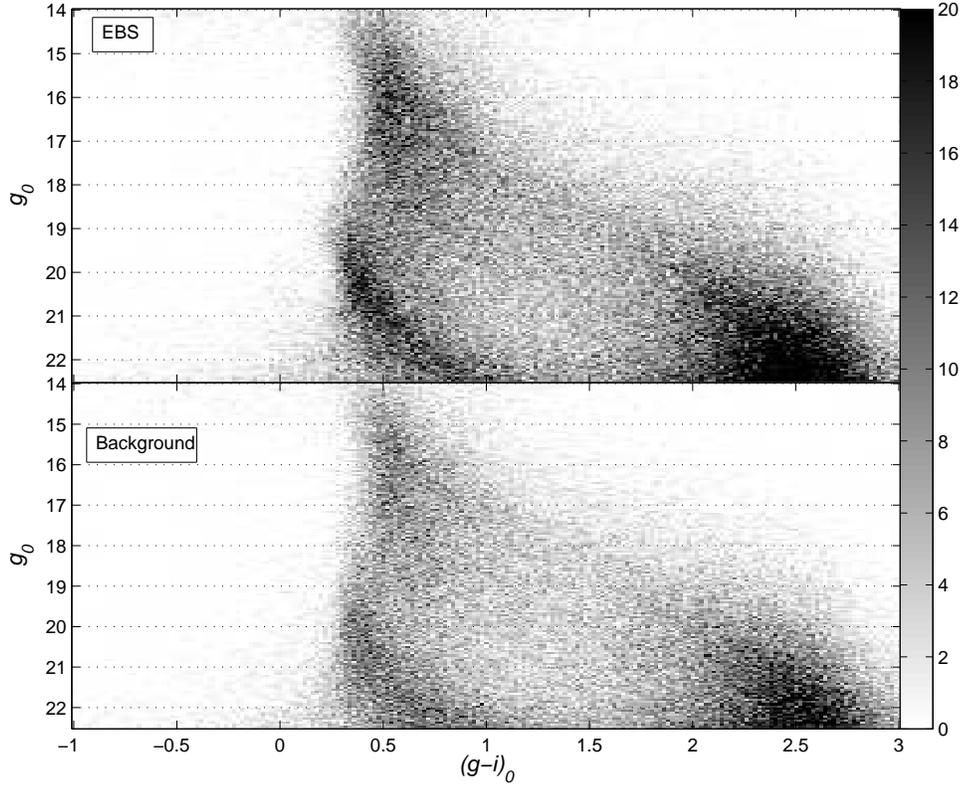}
\caption{Color-magnitude Hess diagrams of stars selected from SDSS DR8 within three regions near the EBS. The upper panel shows stars in the EBS region selected within the lower of the two red boxes in Figure~16; this box is defined by $(-0.15 l + 62.1\arcdeg) < b < (-0.15 l + 64.6\arcdeg)$, $220\arcdeg < l < 230\arcdeg$. A turnoff and main sequence is clearly seen starting at $g_0 \sim 19$ and extending downward -- this is likely due to the EBS. The background region (lower panel) is selected from the same longitude range, but shifted up by $5\arcdeg$ in latitude (as shown by the dashed red box in Figure~16). This box should be far enough away that it will not contain EBS stars. A feature is seen at faint magnitudes in the background region, with highest density at a turnoff magnitude of $g_0 \sim 20.3-20.5$, corresponding to a distance of $\sim16.6-18.2$ kpc, assuming $M_g = 4.2$. This same feature is evident in the EBS region (upper panel), although it is obscured by the much more numerous EBS stars. We suggest that the reason the EBS magnitude peak is so broad in Figure~15 is because the stellar spheroid contributes at $g_0 \gtrsim 19.5$, which skews our Gaussian fit to the peak toward fainter magnitudes.}
\end{figure}

\clearpage
\begin{figure}
\centering
\figurenum{23}
\includegraphics[scale=1.0,angle=0]{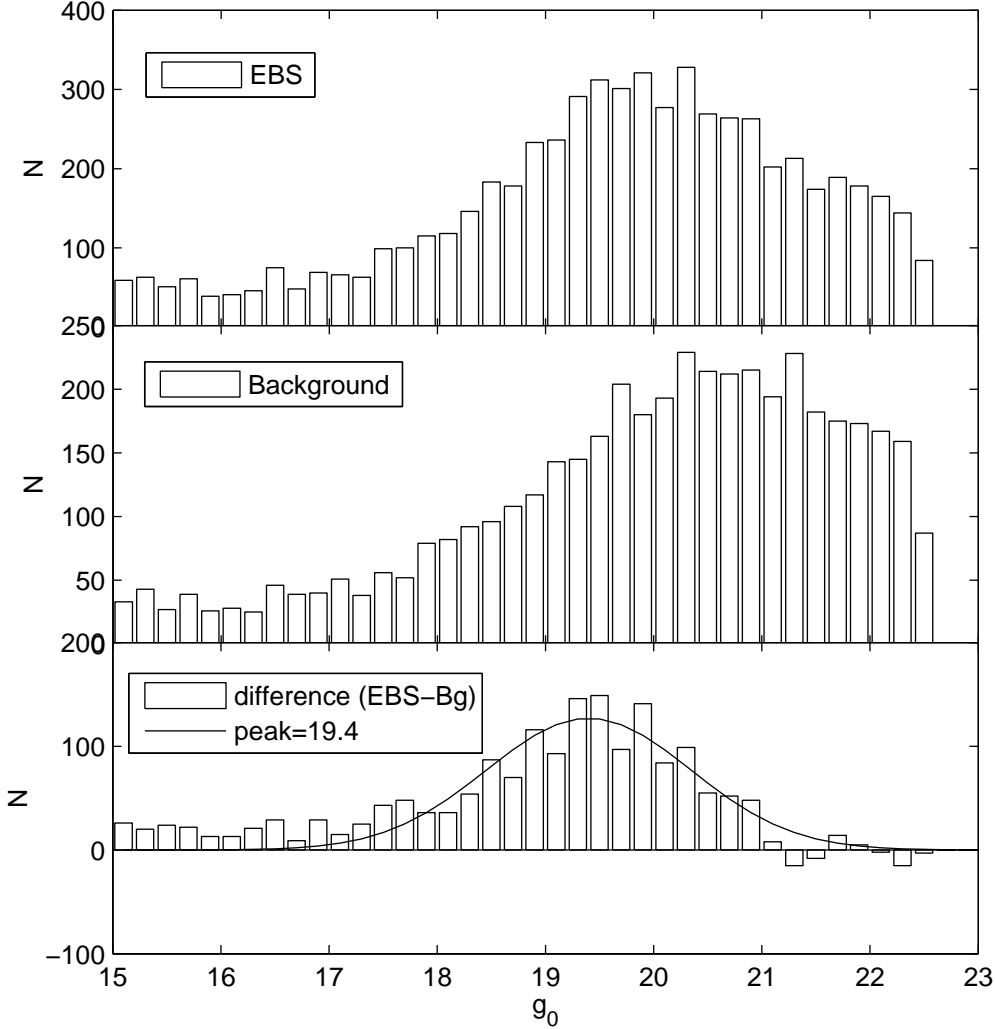}
\caption{Magnitude distribution of F-turnoff stars with $0.2<(g-r)_0<0.3$ and $(u-g)_0>0.4$ in the EBS region (upper panel) and the background comparison region (middle panel) covering an equal-sized area. The lower panel shows the difference obtained by subtracting the background starcounts from those in the EBS region. The contribution from the stellar spheroid at faint magnitudes mostly disappears, leaving a fairly narrow peak that is due to the EBS substructure. We fit the residual in the lower panel with a Gaussian, and find a center at $g_0 = 19.40^{+0.14}_{-0.13}$, which corresponds to a distance to the EBS of $10.96^{+0.73}_{-0.66}$ kpc, assuming $M_g = 4.2$. This shows clearly that spheroid stars were skewing our previous fits (see Figure~15) to larger distances, and that the true answer is more in line with previous estimates of the EBS distance \citep{2011ApJ...738...98G, 2009ApJ...703.2177S}.}
\end{figure}

\end{document}